 \documentclass[final,3p,times,twocolumn]{elsarticle}

\biboptions{numbers,sort&compress}
\usepackage{url}
\usepackage{longtable,rotating,multirow}
\usepackage{setspace}
\usepackage{color}

\usepackage{graphicx}      
\usepackage{natbib}        
\usepackage{amsmath}
\usepackage{amsfonts, amssymb}
\usepackage{bm}
\usepackage{threeparttable}

\usepackage{lineno}

\journal{Journal of Process Control}

\newtheorem{theorem}{Theorem}
\newtheorem{corollary}{Corollary}
\newtheorem{definition}{Definition}
\newtheorem{lemma}{Lemma}

\newtheorem{remark}{Remark}
\begin{document}

\begin{frontmatter}

\title{Detection and Detectability of Intermittent Faults Based on Moving Average $T^2$ Control Charts with Multiple Window Lengths}

\author[Beijing,UMD]{Yinghong~Zhao}\ead{zyh14@mails.tsinghua.edu.cn}
\author[Beijing]{Xiao~He}
\author[UMD]{Michael~G.~Pecht}
\author[Beijing]{Junfeng~Zhang}
\author[Qingdao,Beijing]{Donghua~Zhou\corref{cor}}\ead{zdh@mail.tsinghua.edu.cn}
\address[Beijing]{Department of Automation, Beijing National Research Center for Information Science and Technology (BNRist), Tsinghua University, Beijing 100084, China}
\address[Qingdao]{College of Electrical Engineering and Automation, Shandong University of Science and Technology, Qingdao 266590, China}
\address[UMD]{Center for Advanced Life Cycle Engineering (CALCE), University of Maryland, College Park, MD 20742, USA}
\cortext[cor]{This work was supported by the National Natural Science Foundation of China (NSFC) under Grants 61751307, 61733009, the Research Fund for the Taishan Scholar Project of Shandong Province of China (LZB2015-162), the Key Project from Natural Sciences Foundation of Guangdong Province under Grant 2018B030311054, and the BNRist Program under Grant BNR2019TD01009. Corresponding author: Donghua~Zhou.}

\begin{abstract}
So far, problems of intermittent fault (IF) detection and detectability have not been fully investigated in the multivariate statistics framework. The characteristics of IFs are small magnitudes and short durations, and consequently traditional multivariate statistical methods using only a single observation are no longer effective. Thus in this paper, moving average $T^2$ control charts (MA-TCCs) with multiple window lengths, which simultaneously employ a bank of MA-TCCs with different window lengths, are proposed to address the IF detection problem. Methods to reduce false/missing alarms and infer the IFs' appearing and disappearing time instances are presented. In order to analyze the detection capability for IFs, definitions of guaranteed detectability are introduced, which is an extension and generalization of the original fault detectability concept focused on permanent faults (PFs). Then, necessary and sufficient conditions are derived for the detectability of IFs, which may appear and disappear several times with different magnitudes and durations. Based on these conditions, some optimal properties of two important window lengths are further discussed. In this way, a theoretical framework for the analysis of IFs' detectability is established as well as extended discussions on how the theoretical results can be adapted to real-world applications. Finally, simulation studies on a numerical example and the continuous stirred tank reactor (CSTR) process are carried out to show the effectiveness of the developed methods.
\end{abstract}

\begin{keyword}
Intermittent fault \sep fault detection and detectability \sep moving average $T^2$ control chart \sep multiple window lengths


\end{keyword}

\end{frontmatter}


\section{Introduction}\label{IntroductionSec}
Data-driven fault detection (FD) for large-scale industry processes has received considerable attention over the past decades \cite{Russell2000Data}. Due to its ability to handle high-dimensional and correlated process variables, the multivariate statistical process monitoring (MSPM) methodology is one of the most effective data-driven techniques for FD and process monitoring \cite{Qin2012Survey}.
MSPM uses multivariate control charts such as Hotelling's $T^2$ statistic, principal component analysis (PCA), partial least squares (PLS), independent component analysis (ICA) or hidden Markov model (HMM)-based control charts \cite{Yin2014A}. According to how fault progresses in time, Isermann \cite{Isermann2005Model} has classified fault into three types: abrupt fault, incipient fault and intermittent fault. Both abrupt fault and incipient fault belong to the category of permanent faults (PFs).

With the rapid development of highly complex technologies, intermittent faults (IFs) have become a serious threat to system reliability.
An IF is a kind of non-permanent fault that often recurs due to the same cause and lasts within a limited period of time \cite{Zhou2020Review,Zhao2018Intermittent,Zhao2019Detecting}. IFs are common in a variety of fields \cite{Correcher2012Intermittent,Zhang2019ANovel} and have imposed an enormous financial burden on electronics, satellites and many other industries \cite{Bakhshi2014Intermittent}. Moreover, IFs tend to get worse over time and may eventually become permanent, resulting in the disruption or breakdown of industrial processes. The detection of IFs can effectively reduce the occurrence of catastrophic faults and is an important means to improve system reliability and security. Thus in recent years, IFs have gradually received noticeable interest from both academia and industry \cite{Jiang2003Diagnosis,Allahham2010Monitoring,Carvalho2017Diagnosability,Yan2018Detection,Zhang2019Robust,Zhang2020Intermittent,Zhang2020Robust,Obeid2016Modeling,Obeid2017Early,Zanardelli2007Identification,Singh2012Decision,Cai2017A}, and a review paper aiming to provide an overall picture of historical, current, and future developments in this area has been published \cite{Zhou2020Review}. Problems of detecting IF as well as its detectability in discrete event systems have been addressed in \cite{Jiang2003Diagnosis,Allahham2010Monitoring,Carvalho2017Diagnosability}.
In additional, detection of IFs has been studied for linear stochastic systems \cite{Yan2018Detection,Zhang2019Robust} with parameter uncertainties \cite{Zhang2020Intermittent,Zhang2020Robust}. Note that system models need to be known in these methods. As for data-driven methods, wavelet transform method has been utilized to detect intermittent interturn faults in a synchronous motor \cite{Obeid2016Modeling,Obeid2017Early}. In \cite{Zanardelli2007Identification}, short-time Fourier transform and undecimated discrete wavelet transform have been used to detect intermittent electrical/mechanical faults in motors. In \cite{Singh2012Decision}, the decision forest method has been employed to investigate IFs via feature selection and classification. A dynamic-bayesian-network-based method has been presented to detect IFs in electronic systems \cite{Cai2017A}. These methods require historical data of various faults.

So far, the IF detection (IFD) and detectability problems have not been fully investigated in the MSPM framework, where historical data of faults are not necessary. The characteristics of IFs are small magnitudes and short durations. The magnitude of IF can be as small as incipient fault while its duration is shorter. Thus, IFs are even more difficult to detect than incipient faults.
It has been indicated \cite{Ji2017Incipient} that traditional MSPM methods using only a single observation such as PCA, PLS and ICA are not sensitive to incipient faults, thus not to mention IFs.
Fortunately, several studies \cite{Wold1994Exponentially,Wachs1999Improved,Chen2001Principle} have shown that faults with small magnitudes can be efficiently detected by employing a time window, i.e., the moving average (MA) or moving window (MW) techniques, giving birth to the MA-PCA \cite{Ji2016Incipient}, MA-PLS \cite{Ji2017Incipient}, MW-PCA \cite{Xun2005Process}, MW-HMM \cite{Lin2018Multimode} and so on. This has paved the way for our investigation of the IFD problem.

However, selections of window lengths in these works have not considered the characteristics of fault duration.
Moreover, existing methods have only considered using a single window length. In terms of using multiple window lengths simultaneously, detection and detectability of IFs have not been fully investigated in available literature due to the complexity of integrating varied detection results given by different window lengths. These issues constitute the main motivations of our present study.
Some other important FD methods that also employ a time window are the dynamic MSPM methods, such as dynamic PCA (DPCA), canonical variate analysis (CVA) and stationary/nonstationary-hybrid-characteristics-based dissimilarity analysis \cite{Zhao2018Incipient}. Note that in this paper, process data are assumed to be independent, and thus these methods will turn into traditional single-observation-based MSPM methods which are not sensitive to IFs, or the dissimilarity analysis method \cite{Zhao2007Dissimilarity,Zhao2009Nonlinear}.
As for the dissimilarity analysis, it is an advanced MSPM method that also employs a time window, and has shown a favorable performance for incipient fault detection and isolation \cite{Zhao2017ASparse}. It usually needs a large window length to calculate the covariance matrix of online data set \cite{Pilario2018Canonical}. Considering that the durations of IFs are always limited, the use of dissimilarity-based methods for IFD still requires further justification.

Hotelling's $T^2$ statistic is a well-known function of the likelihood ratio criterion, which consequently makes it admissible and uniformly powerful in certain classes of hypothesis tests \cite{Wierda1994Multivariate}. Thus in this paper, $T^2$ statistic has been combined with the MA technique to constitute a bank of MA $T^2$ control charts (MA-TCCs) with different window lengths.
The main contributions of the present paper are summarized as follows:
1) MA-TCCs with multiple window lengths, including methods to exclude/compensate false/missing alarms and infer the appearing and disappearing time instances of IFs are proposed based on the detectability of each single MA-TCC.
2) The concept of IF detectability is defined for the first time, which is an extension and generalization of the original fault guaranteed detectability concept focused on PFs.
3) A theoretical framework for the analysis of IF detectability is established. Necessary and sufficient conditions for the detectability of IFs, which may appear and disappear several times with different magnitudes and durations are given. Extended discussions on how theoretical results can help detect IFs in practical applications are also presented.

The remainder of this paper is organized as follows. In Section \ref{PreliminariesSec}, the MA-TCC is introduced for the IFD problem. Then the detectability of IFs is analyzed in Section \ref{DetectabilitySec}.
MA-TCCs with multiple window lengths are utilized to reduce false/missing alarms and infer IFs' appearing and disappearing time instances in Section \ref{MATCCsMSec}. Simulation results are presented in Section \ref{SimulationSec}, and conclusions are given in Section \ref{ConclusionSec}.

{\bf Notation:} Bold-face notations in lowercase and uppercase stand for vectors and matrices respectively, so as to distinguish them from scalars. A bold-face notation in [] such as $[\bm{k}]$, is used to highlight the scalar in [].
$\mathbf{A}^T$ and $\mathbf{A}^{-1}$ stand for the transpose and the inverse of a matrix $\mathbf{A}$, respectively.
${\mathbb N_p}(\bm{\mu},\mathbf{\Sigma})$ represents a $p$-dimensional normal distribution with expectation $\bm{\mu}$ and covariance matrix $\mathbf{\Sigma}$.
${\mathbb W_p}(N,\mathbf{\Sigma})$ represents a $p$-dimensional Wishart distribution with $N$ degrees of freedom.
${\mathbb F}(p,N-p)$ is a central $F$ distribution with $p$ and $N-p$ degrees of freedom.
${\mathbb F}_{\alpha}(p,N-p)$ is the $1-\alpha$ percentile of the central $F$ distribution with $p$ and $N-p$ degrees of freedom.
${\mathbb N}_+$ and ${\mathbb R}_+$ are the sets of positive integers and positive real numbers, respectively.
$[x]^+$ is the minimum integer no less than $x$, and $[x]^-$ is the maximum integer no more than $x$. $x\!>\!(\geq)\max\{y,z\}$ means if $y\geq z$, then $x>y$, otherwise $x\geq z$. $\emptyset$ is the empty set, and $\left[a,b\right)=\{x\in{\mathbb R}:a\leq x<b\}$. $\triangleq$ is to give a definition.

\section{Preliminaries}\label{PreliminariesSec}
\subsection{Hotelling's $T^2$ distribution}\label{HotellingSub}
The following lemma is the key result regarding Hotelling's $T^2$ distribution, see \cite{Anderson2003An}.
\begin{lemma}\label{LemHotellingT}
Let $T^2=\mathbf{x}^T\mathbf{S}^{-1}\mathbf{x}$, where $\mathbf{x}$ and $\mathbf{S}$ are independently distributed with $\mathbf{x}\sim{\mathbb N}_p(\bm{\mu},\mathbf{\Sigma})$ and $N\mathbf{S}\sim{\mathbb W}_p(N,\mathbf{\Sigma})$, where $N\geq{p}$. Then
\begin{align}
T^2\sim \frac{Np}{N-p+1}{\mathbb F}(p,N-p+1;\epsilon^2),
\end{align}
where the noncentrality parameter $\epsilon^2=\bm{\mu}^T\mathbf{\Sigma}^{-1}\bm{\mu}$.
\end{lemma}

\subsection{Moving average $T^2$ control chart (MA-TCC)}\label{MATCCSub}
Suppose we have collected $N$ independent samples $\mathbf{x}_1,\mathbf{x}_2,\cdots,\mathbf{x}_N$ from ${\mathbb N}_p(\bm{\mu},\mathbf{\Sigma})$ under certain sampling rate as training data, which can represent the statistic characteristics of systems' normal conditions. We also collect current process data with a same sampling rate.
Then the MA-TCC concerns the analysis of latest $W$ new current process data $\mathbf{x}^f_{k-W+1},\cdots,\mathbf{x}^f_{k-1},\mathbf{x}^f_k$ at each time $k$, to determine whether the process is statistically fault-free or not.
We ordinarily assume that current process data are independent, and are identically distributed with the training data except for a different mean $\bm{\mu}_f$.
Thus, MA-TCC is transformed into a hypothesis testing ${\mathcal H}_0: \bm{\mu}_f=\bm{\mu}$ versus ${\mathcal H}_1: \bm{\mu}_f\neq\bm{\mu}$.
In practice, parameters $\bm{\mu}, \bm{\mu}_f, \mathbf{\Sigma}$ are unknown, and we only know the sample means $\bar{\mathbf{x}}, \bar{\mathbf{x}}^f_k$ and the sample covariance matrix $\mathbf{S}$ instead:
\begin{align}
\bar{\mathbf{x}}^f_k&=\frac{1}{W}\sum\limits_{i=1}^{W}\mathbf{x}^f_{k-W+i},\; \bar{\mathbf{x}}=\frac{1}{N}\sum\limits_{i=1}^{N}\mathbf{x}_i,\nonumber\\
\mathbf{S}&=\frac{1}{N-1}\sum\limits_{i=1}^{N}(\mathbf{x}_i-\bar{\mathbf{x}})(\mathbf{x}_i-\bar{\mathbf{x}})^T.
\end{align}
We also know that the sample means $\bar{\mathbf{x}}, \bar{\mathbf{x}}^f_k$ and the sample covariance matrix $\mathbf{S}$ are independently distributed with
\begin{align}\label{SampleMeanCov}
(\bar{\mathbf{x}}^f_k-\bar{\mathbf{x}})&\sim{\mathbb N}_p(\bm{\mu}_f-\bm{\mu},\frac{N+W}{NW}\mathbf{\Sigma}),\nonumber\\
(N-1)\mathbf{S}&\sim{\mathbb W}_p(N-1,\mathbf{\Sigma}).
\end{align}
Then under normal conditions, the MA-TCC with window length $W$, denoted as MA-TCC($W$), at time instance $k$ is (see Appendix A)
\begin{align}\label{T2Conse}
T^2_k(W)&=(\bar{\mathbf{x}}^f_k-\bar{\mathbf{x}})^T\mathbf{S}^{-1}(\bar{\mathbf{x}}^f_k-\bar{\mathbf{x}})\nonumber\\
&\sim\frac{p(N+W)(N-1)}{NW(N-p)}{\mathbb F}(p,N-p).
\end{align}
For a given significance level $\alpha$, the process is considered normal, i.e., to accept ${\mathcal H}_0: \bm{\mu}_f=\bm{\mu}$, if
\begin{align}\label{UclConsecutive}
T^2_k(W)\leq\delta^2_W=\frac{p(N+W)(N-1)}{NW(N-p)}{\mathbb F}_{\alpha}(p,N-p),
\end{align}
where $\delta^2_W$ is the control limit of the MA-TCC($W$). Inequality \eqref{UclConsecutive} gives the acceptance region of the hypothesis testing. For further reference, denote $\delta^2\!=\!\delta^2_1$ as the control limit of the TCC with a single observation.

\begin{remark}
By introducing a time window, the control limit of the MA-TCC($W$) decreases with the increase of $W$ as follows
\begin{align}\label{UclRelation}
\delta^2_W=\frac{N+W}{W(N+1)}\delta^2\xrightarrow{W\to\infty}\frac{1}{N+1}\delta^2.
\end{align}
Note that, according to (\ref{UclRelation}), $\delta^2_W$ cannot decrease to $0$ even though $W\to\infty$. This is because the chosen $\delta^2_W$ is based on a significance level $\alpha$, which means that $\delta^2_W$ should always guarantee the probability of the type I error (i.e., false alarm rate in this specific FD application) in this hypothesis testing less than $\alpha$. Since the number of training samples $N$ is finite, estimation error for the presumed normal distribution ${\mathbb N}_p(\bm{\mu},\mathbf{\Sigma})$ is inevitable and thus must be tolerated by the choice of $\delta^2_W$.
On the other hand, the estimation error converges to zero when $N$ is infinite, and then $\delta^2_W$ can decrease to $0$. This can be verified by $\delta^2_\infty\to{0}$ when $N\to\infty$.
\end{remark}

\section{Detectability analysis}\label{DetectabilitySec}
\subsection{Definitions of guaranteed detectability}
From both an analytical and a practical point of view, it is important to know whether a fault is detectable by the proposed methods.
Consider the following widely used fault model in the MSPM framework \cite{Qin2003Statistical,Alcala2009Reconstruction}
\begin{align}\label{FaultModel}
\mathbf{x}^f_k=\mathbf{x}^{*}_k+\mathbf{\Xi}_{k}\mathbf{f}_k,
\end{align}
where $\mathbf{x}^*_k$ represents the process fluctuation under normal conditions, $\mathbf{\Xi}_{k}$ is the direction of the fault in time instance $k$, and $\|\mathbf{f}_k\|$ is its magnitude.
The fault-free part $\mathbf{x}^*_k$ usually represents a normal steady-state condition. In this way, the above fault model represents a mismatch between actual measurements and normal process fluctuations in the event of a fault.
By introducing the time window, we have
\begin{align}\label{FaultModelWin}
\bar{\mathbf{x}}^f_k=\bar{\mathbf{x}}^{*}_k+\bar{\mathbf{\Xi}}_{k}\bar{\mathbf{f}}_k,\; \bar{\mathbf{x}}^*_k&=\frac{1}{W}\sum\limits_{i=1}^{W}\mathbf{x}^*_{k-W+i},
\end{align}
where $\bar{\mathbf{\Xi}}_k\bar{\mathbf{f}}_k$ is the effect of all faults in the time window.
Since the fault-free process has been assumed to follow a normal distribution, we have $\bar{\mathbf{x}}^*_k\sim{\mathbb N}_p(\bm{\mu},\frac{1}{W}\mathbf{\Sigma})$.
To analyze the fault detectability of MA-TCC($W$), we introduce the following condition:
\begin{align}\label{AssuAcceptRegion}
\|\mathbf{S}^{-1/2}(\bar{\mathbf{x}}^{*}_{k}-\bar{\mathbf{x}})\|^2\leq\delta^2_W.
\end{align}
\begin{remark}
Since fault-free data come from a normal distribution, then for any fault, there is no guarantee (with 100\% probability) to detect it. Thus, we need an additional condition, under which the fault detectability can be defined. One way is to employ another significance level $\beta$, and define the detectability under the condition that the type II error (i.e., missing alarm rate) is less than $\beta$. Another way is to employ the condition \eqref{AssuAcceptRegion}, and define the guaranteed detectability on the basis of it. This way has been widely accepted by literature addressing fault detectability problems in the MSPM framework \cite{Qin2003Statistical,Alcala2009Reconstruction,Mnassri2013Generalization,Ji2017Incipient}.
The condition means that the fault-free process $\bar{\mathbf{x}}^*_k$ fluctuates within its acceptance region \eqref{UclConsecutive}. It is commonly employed because it holds with high probability (note $\alpha$ is always small), and furthermore its violation is the underlying cause of false/missing alarms. This mechanism can help us exclude/compensate false/missing alarms in Section \ref{MATCCsMSec}.
\end{remark}

\begin{figure}
\begin{center}
\includegraphics[width=8.4cm]{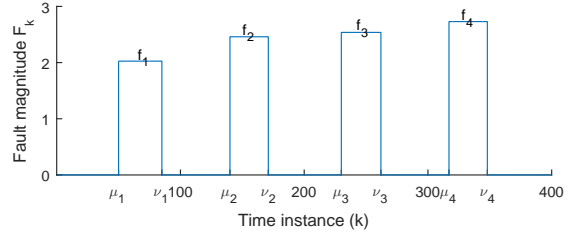}    
\caption{An example of intermittent faults.}
\label{IFexample}
\end{center}
\end{figure}

In the case of IFs, as shown in Fig.~\ref{IFexample}, the corresponding fault model can be represented \cite{Zhou2020Review,Yan2018Detection} by
\begin{align}\label{IFModel}
\mathbf{\Xi}_{k}\mathbf{f}_k=\sum\limits_{q=1}^{\infty}\left[\Gamma(k-\mu_{q})-\Gamma(k-\nu_{q})\right]\bm{\xi}_q{f_q},
\end{align}
where $\Gamma(\cdot)$ is the step function. $\mu_{q}$ and $\nu_{q}$ are the appearing and disappearing time instances of the $q$th IF, respectively. They satisfy $\mu_q\!<\!\nu_q\!<\!\mu_{q+1}\in{\mathbb N}_+$. $\bm{\xi}_q\in{\mathbb R}^{p}$ and $f_q\!>\!0\in{\mathbb R}^{1}$ represent the direction and magnitude of the $q$th IF, respectively. Since the fault duration is limited, IFs do not change much within each active period. Thus, here it is assumed that the fault direction and magnitude within each IF are constant. We do not assume that the fault directions or the fault magnitudes for different active periods are the same, because IFs tend to get worse over time.
Denote the active and inactive duration of the $q$th IF as $\tau^o_q\!=\!\nu_q\!-\!\mu_q$ and $\tau^r_q\!=\!\mu_{q+1}\!-\!\nu_q$, respectively. Note that they are counted by sampling intervals here. Although $\mu_{q},\nu_{q},\tau^o_q,\tau^r_q$ may not be integers, their fractional parts no longer affect the measurements according to \eqref{FaultModel} and \eqref{IFModel}. Thus, for FD we can reset them as $[\mu_{q}]^+,[\nu_{q}]^-,[\tau^o_q]^-,[\tau^r_q]^-$ respectively, and they are integers hereinafter.
Then, the $q$th IF can be represented with five parameters, i.e., $IF(\bm{\xi}_q, f_q, \tau^r_{q-1}, \tau^o_q, \tau^r_q)$.
Denote $\mathcal{P}_{q}=\left\{(\mathcal{P}^o_{q},\mathcal{P}^r_{q})\right\}$ as the set of all possible parameter values of the $q$th IF, where
\begin{align*}
\mathcal{P}^o_{q}&=\left\{(\bm{\xi}_q, f_q, \tau^o_q) \;|\; \bm{\xi}_q\!\in\!{\mathbb R}^{p}, f_q\!\in\!{\mathbb R}_+, \tau^o_q\!\in\!{\mathbb N}_+\right\},\\
\mathcal{P}^r_{q}&=\left\{(\tau^r_q) \;|\; \tau^r_q\!\in\!{\mathbb N}_+\right\}.
\end{align*}
Denote $\mathcal{P}^{q}$ as the set of all possible parameter values of the first $q$ IFs, i.e., $\mathcal{P}^{q}=\left\{ (\mathcal{P}_1,\cdots,\mathcal{P}_{q-1},\mathcal{P}^o_{q}) \right\}$.

The concept of fault detectability in the MSPM framework was first defined in \cite{Dunia1998A,Dunia1998Subspace}, and has been widely adopted to study the detection capability of a variety of MSPM methods \cite{Qin2003Statistical,Alcala2009Reconstruction,Qin2012Survey,Mnassri2013Generalization,Ji2017Incipient}. The defined guaranteed detectability means that the fault is guaranteed to be detected under the condition \eqref{AssuAcceptRegion}. However, the concept has been limited to the case of PFs, which makes fault detectability analyses for IFs impossible.
Compared with the PF detection (PFD) task, additional requirements for IFD \cite{Zhou2020Review,Yan2018Detection} are to determine each appearance (disappearance) of the IF before its subsequent disappearance (appearance), otherwise missing or false alarms occur. Based on these considerations, this paper provides an extension and generalization of the original fault detectability concept \cite{Dunia1998A} to make it suitable for both PFs and IFs.

\begin{definition}\label{DefnIFDrec}
For a given significance level $\alpha$, the disappearance of the $q$th IF is said to be \textbf{guaranteed detectable} by the MA-TCC($W$), if there exists a time instance $\nu_q\!\leq\![\bm{k^{\#}}]\!<\!\mu_{q+1}$ such that for each $k^{\#}\leq[\bm{k}]<\mu_{q+1}$, the detection statistic $T^2_k(W)\!\leq\!\delta^2_W$ is guaranteed for all values of $\bar{\mathbf{x}}^*_k$ in \eqref{AssuAcceptRegion} and for all elements of ${\mathcal P}^{q}$. Note that all $k^{\#}$ constitute a set $\mathcal{K^{\#}}$. Then the MA-TCC($W$)'s alarm delay for the $q$th disappearance is defined as $\nu^d_q(W)\triangleq\inf_{k^{\#}\in\mathcal{K^{\#}}} \left({k^{\#}}-\nu_{q}\right)$. We further denote $k^{\#}_q(W)\triangleq\arg\inf_{k^{\#}\in\mathcal{K^{\#}}}\left({k^{\#}}-\nu_{q}\right)=\nu_{q}+\nu^d_q(W)$.
\end{definition}

\begin{definition}\label{DefnIFDocc}
For a given significance level $\alpha$, the appearance of the $q$th IF is said to be \textbf{guaranteed detectable} by the MA-TCC($W$), if the disappearance of the $(q\!-\!1)$th IF is guaranteed detectable, and there exists a time instance $\mu_{q}\!\leq\![\bm{k^*}]\!<\!\nu_{q}$ such that for each ${k^*}\!\leq\![\bm{k}]\!<\!\nu_{q}$, the detection statistic $T^2_k(W)\!>\!\delta^2_W$ is guaranteed for all values of $\bar{\mathbf{x}}^*_k$ in \eqref{AssuAcceptRegion}. Note that all $k^*$ constitute a set $\mathcal{K^*}$. Then the MA-TCC($W$)'s alarm delay for the $q$th appearance is defined as $\mu^d_q(W)\triangleq\inf_{k^*\in\mathcal{K^*}} \left({k^*}-\mu_{q}\right)$. We further denote $k^*_q(W)\triangleq\arg\inf_{k^*\in\mathcal{K^*}}\left({k^*}-\mu_{q}\right)=\mu_{q}+\mu^d_q(W)$.
\end{definition}

\begin{definition}\label{DefnIFD}
For a given significance level $\alpha$, the $q$th IF is said to be \textbf{guaranteed detectable} by the MA-TCC($W$), if both the appearance and disappearance of the $q$th IF are guaranteed detectable.
\end{definition}

\subsection{Detectability conditions}
Now, we are interested in finding necessary and sufficient conditions for the detectability of each IF.
However, this is a daunting task since detectability of each IF is coupled with previous IFs implicitly and varies with different window lengths. In other words, because of the time window, the $T^2$ statistic is not only influenced by the present IF but may also be influenced by previous IFs, which have different fault directions and magnitudes, and active and inactive durations.
Nevertheless, through the analyses for fault disappearance detectability, the valid range of window lengths can be obtained in Lemma \ref{LemIFDrec}.
Then the detectability of fault appearance with a certain window length is given in Lemmas \ref{LemIFDocc1} and \ref{LemIFDocc2}.
Based on them, the necessary and sufficient conditions for the detectability of each IF with and without the limit of certain window length are finally obtained in Theorems \ref{ThmIFD0}, \ref{ThmIFD}, and \ref{ThmIFD2}.

\begin{lemma}\label{LemIFDrec}
For the MA-TCC($W$) and a given significance level $\alpha$, the disappearance of the $q$th IF is \textbf{guaranteed detectable} if and only if
\begin{align}\label{EquIFDrec}
W\leq\tau^r_{q}.
\end{align}
\end{lemma}
\textbf{Proof.} See Appendix A. \qed

As for the detectability of the $q$th IF's appearance, it follows from Definition \ref{DefnIFDocc} and Lemma \ref{LemIFDrec} that the window length should satisfy $W\leq\tau^r_{q-1}$.
This set can be divided twofold as follows
\begin{align*}
\left\{W\leq\tau^r_{q-1}\right\}=\left\{W\leq\min\{\tau^r_{q-1},\tau^o_q\}\right\}\cup\left\{\tau^o_q<W\leq\tau^r_{q-1}\right\}.
\end{align*}
Then we have the following two lemmas.
\begin{lemma}\label{LemIFDocc1}
For the MA-TCC($W$) and a given significance level $\alpha$, when $W\leq\min\{\tau^r_{q-1},\tau^o_q\}$, the appearance of the $q$th IF is \textbf{guaranteed detectable} if and only if
\begin{align}\label{EquIFDocc1}
\|\mathbf{S}^{-\frac{1}{2}}\bm{\xi}_q{f_q}\|>2\delta_W=2\delta\sqrt{\frac{N+W}{W(N+1)}}.
\end{align}
\end{lemma}
\textbf{Proof.} See Appendix A. \qed

\begin{lemma}\label{LemIFDocc2}
For the MA-TCC($W$) and a given significance level $\alpha$, when $\tau^o_q<W\leq\tau^r_{q-1}$, the appearance of the $q$th IF is \textbf{guaranteed detectable} if and only if
\begin{align}\label{EquIFDocc2}
\|\mathbf{S}^{-\frac{1}{2}}\bm{\xi}_q{f_q}\|\frac{\tau^o_q}{W}>2\delta_W=2\delta\sqrt{\frac{N+W}{W(N+1)}}.
\end{align}
\end{lemma}
\textbf{Proof.} See Appendix A. \qed

\begin{theorem}\label{ThmIFD0}
For the MA-TCC($W$) and a given significance level $\alpha$, the $q$th IF is \textbf{guaranteed detectable} if and only if:

(i) $W\leq\min\{\tau^r_{q-1},\tau^r_{q}\}$; and

(ii) when $W\leq\min\{\tau^r_{q-1},\tau^o_q\}$, inequality (\ref{EquIFDocc1}) holds or when $\tau^o_q<W\leq\tau^r_{q-1}$, inequality (\ref{EquIFDocc2}) holds.
\end{theorem}
\textbf{Proof.} Directly derived from Lemmas \ref{LemIFDrec}, \ref{LemIFDocc1}, and \ref{LemIFDocc2}. \qed

\begin{theorem}\label{ThmIFD}
For the MA-TCC and a given significance level $\alpha$, the $q$th IF is \textbf{guaranteed detectable} if and only if
\begin{align}\label{EquIFD}
\frac{N+1}{4\delta^2}\|\mathbf{S}^{-\frac{1}{2}}\bm{\xi}_q{f_q}\|^2-1>\frac{N}{\min\{\tau^r_{q-1},\tau^o_q,\tau^r_{q}\}}.
\end{align}
Then the window length $W$ can be chosen such that
\begin{align}\label{WinLengthIFD}
&\frac{1}{N}\left(\frac{N+1}{4\delta^2}\|\mathbf{S}^{-\frac{1}{2}}\bm{\xi}_q{f_q}\|^2-1\right)>\frac{1}{W}>(\geq)\\
&\max\!\left\{ \frac{ \frac{N}{2\tau^o_q}\!+\!\sqrt{\left(\frac{N}{2\tau^o_q}\right)^2\!+\!\frac{N+1}{4\delta^2}\|\mathbf{S}^{-\frac{1}{2}}\bm{\xi}_q{f_q}\|^2} }
{\tau^o_q\frac{N+1}{4\delta^2}\|\mathbf{S}^{-\frac{1}{2}}\bm{\xi}_q{f_q}\|^2},\frac{1}{\min\{\tau^r_{q-1},\tau^r_{q}\}} \right\},\nonumber
\end{align}
and the corresponding alarm delays are
\begin{align}\label{DelayIFD}
\mu^d_q(W)&=\left[\! \sqrt{\frac{W(N\!+\!W)}{N+1}}\frac{2\delta}{\|\mathbf{S}^{-\frac{1}{2}}\bm{\xi}_q{f_q}\|} \right]^+\!-\!1\leq\min\left\{W,\tau^o_q\right\}-\!1,\nonumber\\
\nu^d_q(W)&=W\!-\!1\leq\tau^r_{q}\!-\!1,
\end{align}
which are all increasing functions of $W$.
\end{theorem}
\textbf{Proof.} See Appendix A. \qed

\begin{remark}\label{RmkLowSamplingRate}
Theorem \ref{ThmIFD} means that, for the $q$th IF with a fixed direction $\bm{\xi}_q$ and magnitude $f_q$, if its duration is less than a certain number of sampling interval, i.e. $W^*$ given by \eqref{EquOptWDirMag}, then it is not guaranteed detectable. Now, we fix the direction, magnitude and duration of the $q$th IF. It can be seen that if the sampling rate keeps decreasing, namely, the sampling interval keeps increasing, then the number of sampling interval that the $q$th IF lasts keeps decreasing. If this number decreases below $W^*$, then the $q$th IF is not guaranteed detectable due to sporadic data samples.
\end{remark}

\begin{theorem}\label{ThmIFD2}
For the MA-TCC and a given significance level $\alpha$, the $q$th IF is \textbf{guaranteed detectable} if and only if $W\!=\!\min\{\tau^r_{q-1},\tau^o_q,\tau^r_{q}\}$ satisfies inequality (\ref{WinLengthIFD}).
\end{theorem}
\textbf{Proof.} It follows from the proof of Theorem \ref{ThmIFD} that if the $q$th IF is guaranteed detectable, then (\ref{EquIFD}) holds and consequently $W\!=\!\min\{\tau^r_{q-1},\tau^o_q,\tau^r_{q}\}$ satisfies inequality (\ref{WinLengthIFD}).
On the other hand, if $W=\min\{\tau^r_{q-1},\tau^o_q,\tau^r_{q}\}$ is one of the solutions of inequality (\ref{WinLengthIFD}), then we derive (\ref{EquIFD}) directly and the proof of Theorem \ref{ThmIFD2} is complete. \qed

\begin{remark}
Now we have obtained three necessary and sufficient conditions for IFs to be guaranteed detectable. For the MA-TCC, Theorem \ref{ThmIFD0} gives a necessary and sufficient condition for the detectability of each IF considering certain window length.
Theorem \ref{ThmIFD} gets more general results by investigating detectability conditions without the limit of a certain window length. In Theorem \ref{ThmIFD2}, the necessary and sufficient condition is given by solving an inequality that the window length satisfies.
Note that the constraints $\nu_q\!\leq\!{k^{\#}}$ and $\mu_{q}\!\leq\!{k^*}$ in Definitions \ref{DefnIFDrec} and \ref{DefnIFDocc} are optional, and have no influence on the subsequent detectability conditions, since Lemmas \ref{LemIFDrec}-\ref{LemIFDocc2} and Theorems \ref{ThmIFD0}-\ref{ThmIFD2} still hold if we remove the constraints from their proofs.
\end{remark}

In practice, we may not know exactly the parameter information of IFs, but lower bounds of fault parameters are relatively easy to be obtained by analyzing historical data or operational conditions. Then we have the following corollary according to Theorems \ref{ThmIFD} and \ref{ThmIFD2}.
\begin{corollary}\label{ColIFDlow}
For the MA-TCC and a given significance level $\alpha$, the $q$th IF is \textbf{guaranteed detectable} if $IF(\bm{\xi}_q,\tilde{f}_q,\tilde{\tau}^r_{q-1},\tilde{\tau}^o_q,\tilde{\tau}^r_{q})$ is guaranteed detectable, where $\tilde{f}_q,\tilde{\tau}^r_{q-1},\tilde{\tau}^o_q,\tilde{\tau}^r_{q}$ are lower bounds of $f_q,\tau^r_{q-1},\tau^o_q,\tau^r_{q}$ respectively.
Then the window length $W$ can be chosen such that
\begin{align}\label{WinLengthIFDlow}
&\frac{1}{N}\left(\frac{N+1}{4\delta^2}\|\mathbf{S}^{-\frac{1}{2}}\bm{\xi}_q{\tilde{f}_q}\|^2-1\right)>\frac{1}{W}>(\geq)\\
&\max\!\left\{
\frac{ \frac{N}{2\tilde{\tau}^o_q}\!+\!\sqrt{\left(\frac{N}{2\tilde{\tau}^o_q}\right)^2\!+\!\frac{N+1}{4\delta^2}\|\mathbf{S}^{-\frac{1}{2}}\bm{\xi}_q{\tilde{f}_q}\|^2} }
{\tilde{\tau}^o_q\frac{N+1}{4\delta^2}\|\mathbf{S}^{-\frac{1}{2}}\bm{\xi}_q{\tilde{f}_q}\|^2},\frac{1}{\min\{\tilde{\tau}^r_{q-1},\tilde{\tau}^r_{q}\}} \right\},\nonumber
\end{align}
and the corresponding alarm delays are
\begin{align}
&\mu^d_q(W)\leq\left[ \sqrt{\frac{W(N+W)}{N+1}}\frac{2\delta}{\|\mathbf{S}^{-\frac{1}{2}}\bm{\xi}_q{\tilde{f}_q}\|} \right]^+\!-\!1
\triangleq\tilde{\mu}^d_q(W)\leq\tilde{\tau}^o_q\!-\!1,\nonumber\\
&\nu^d_q(W)=W\!-\!1\triangleq\tilde{\nu}^d_q(W)\leq\tilde{\tau}^r_{q}\!-\!1.
\end{align}
\end{corollary}
\textbf{Proof.} Directly derived from Theorem \ref{ThmIFD}. \qed

Remark \ref{RmkLowSamplingRate} has demonstrated that, to avoid the loss of detectability, the sampling rate should not be too low. However, it has been pointed out that sampling as fast as possible is also inadvisable for FD \cite{Russell2000Data}. Since the time constant is closely related to the dynamic characteristics of a process, it can help us determine a proper sampling rate at which the primary process characteristics are captured. A practical rule is to choose the sampling interval as one-tenth of the time constant \cite{Russell2000Data}. Then, if at this sampling rate, the detectability condition is satisfied, we can accept it. Otherwise, we suggest reasonably shortening the sampling interval to improve the detectability, though this may sacrifice some detection performance such as causing more false alarms.

A PF can be viewed as an IF with infinite active duration. Thus, all the above analyses are applicable to PF when we set $\tau^r_{q-1},\tau^o_q,\tau^r_{q}\to\infty$.
For a PF with fault direction $\bm{\xi}_{PF}$ and magnitude $f_{PF}$, the detectability conditions are given by the following theorem.

\begin{corollary}\label{ColPFD}
For the MA-TCC and a given significance level $\alpha$, a PF is \textbf{guaranteed detectable} if and only if
\begin{align}\label{EquPFD}
\frac{1}{N}\left(\frac{N+1}{4\delta^2}\|\mathbf{S}^{-\frac{1}{2}}\bm{\xi}_{PF}f_{PF}\|^2-1\right)>0.
\end{align}
Then the window length $W$ can be chosen such that
\begin{align}\label{WinLengthPFD}
\frac{1}{N}\left(\frac{N+1}{4\delta^2}\|\mathbf{S}^{-\frac{1}{2}}\bm{\xi}_{PF}f_{PF}\|^2-1\right)>\frac{1}{W}>0,
\end{align}
and the corresponding alarm delay is
\begin{align}\label{DelayPFD}
\mu^d_{PF}(W)=\left[ \sqrt{\frac{W(N+W)}{N+1}}\frac{2\delta}{\|\mathbf{S}^{-\frac{1}{2}}\bm{\xi}_{PF}f_{PF}\|} \right]^+-1.
\end{align}
\end{corollary}
\textbf{Proof.} Directly derived from Theorem \ref{ThmIFD}. \qed

\begin{remark}
It can be easily seen that the detection of IFs is much harder than PFs, since the detectability condition for IFs (\ref{EquIFD}) is stricter than it is for PFs (\ref{EquPFD}).
\end{remark}

\subsection{Two important window lengths}
The above theorems and corollaries give the detectability conditions and selection criteria of window length according to the knowledge of fault parameters, which are effective for targeted IFs with prior knowledge.
Note that in order to implement the above theorems and corollaries, we need to know at least five fault parameters, i.e., $\{\bm{\xi}_q,f_q,\tau^r_{q-1},\tau^o_q,\tau^r_{q}\}$, or their lower bounds instead.
In practice, we may not have all the parameter information of IFs, but more likely know part of the fault information. The following theorems study how to choose a proper $W$ in these cases.

In the following, we consider the set of $q$th IFs with same $(\bm{\xi}_q,{f_q})$ as well as different active and inactive durations, and further denote it as $IF(\bm{\xi}_q, f_q)$. Note that $IF(\bm{\xi}_q, f_q, *, *, *)$ is an element in the set $IF(\bm{\xi}_q, f_q)$.
\begin{theorem}\label{OptWDirMag}
For the set of $IF(\bm{\xi}_q, f_q)$ and a given significance level $\alpha$, the MA-TCC($W^*$)
\begin{align}\label{EquOptWDirMag}
W^*=\left[ \frac{N}{\frac{N+1}{4\delta^2}\|\mathbf{S}^{-\frac{1}{2}}\bm{\xi}_q{f_q}\|^2-1} \right]^+,
\end{align}
has the following properties:

(i) for $\forall IF(\bm{\xi}_q, f_q, \tau^r_{q-1}, \tau^o_q, \tau^r_q)\in IF(\bm{\xi}_q, f_q)$, if it is not guaranteed detectable with $W^*$, then there is no other $W$ making it guaranteed detectable;

(ii) for $\forall IF(\bm{\xi}_q, f_q, \tau^r_{q-1}, \tau^o_q, \tau^r_q)\in IF(\bm{\xi}_q, f_q)$, if it is guaranteed detectable with $W^*$, then there is no other $W$ making its corresponding alarm delay smaller.
\end{theorem}
\textbf{Proof.} See Appendix A. \qed

Likewise, we consider the set of $q$th IFs with same $(\tau^o_q, \tau^r_q)$ as well as different fault direction and magnitude, and further denote it as $IF(\tau^o_q, \tau^r_q)$.
\begin{theorem}\label{OptWDur}
For the set of $IF(\tau^o_q, \tau^r_q)$ and a given significance level $\alpha$, the MA-TCC($W^\#$)
\begin{align}\label{EquOptWDur}
W^\#=\min\{\tau^r_{q-1},\tau^o_q,\tau^r_{q}\},
\end{align}
has the following properties:

(i) for $\forall IF(\bm{\xi}_q, f_q, \tau^r_{q-1}, \tau^o_q, \tau^r_q)\in IF(\tau^o_q, \tau^r_q)$, if it is not guaranteed detectable with $W^\#$, then there is no other $W$ making it guaranteed detectable.
\end{theorem}
\textbf{Proof.} Similar to Theorem \ref{OptWDirMag}.

\begin{remark}
The above theorems study the optimal choice of window length when only parts of the fault information are available.
Theorem \ref{OptWDirMag} illustrates that given the fault direction and magnitude information, the optimal choice of window length is (\ref{EquOptWDirMag}) in consideration of both detection capability and alarm delay. This is quite useful for cases where the maximum tolerable deviation is known.
Theorem \ref{OptWDur} illustrates that in order to guarantee the detectability of smaller IFs, the proper window length is (\ref{EquOptWDur}).
\end{remark}

\section{IFD based on MA-TCCs with multiple window lengths}\label{MATCCsMSec}
The advantage and disadvantage of introducing the time window for IFD are apparent, i.e., the improved sensitivity and the introduced alarm delay. Thus, it is natural to consider MA-TCCs with multiple window lengths, denoted as MA-TCCs(M), for the IFD problem.
However, detection results given by different window lengths are often inconsistent due to false or missing alarms in real-world applications.
Thus, methods to exclude false alarms and compensate missing alarms are proposed first on the basis of the detectability analyses presented in the previous section. Then, IFs' appearing and disappearing time instances $\mu_{q}$,$\nu_{q}$ are determined by integrating different MA-TCCs' inference results.

\subsection{False alarms exclusion and missing alarms compensation}
\begin{figure}
\begin{center}
\includegraphics[width=8.4cm]{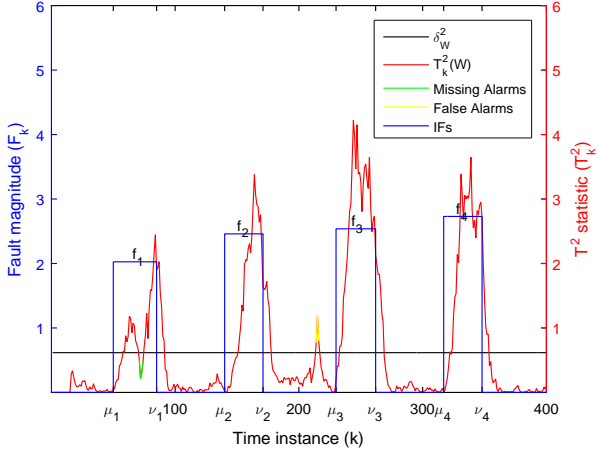}    
\caption{An example of false and missing alarms.}
\label{MFalarm}
\end{center}
\end{figure}

When IFs are guaranteed detectable by the MA-TCC($W$), the underlying cause of false/missing alarms is the violation of condition \eqref{AssuAcceptRegion}. When no fault occurs, it leads to false alarms if $\bar{\mathbf{x}}^*_k$ goes beyond its acceptance region, as shown in Fig.~\ref{MFalarm} with yellow line. When IF occurs, it may lead to reduction of the IF's impact (consequently missing alarms) if $\bar{\mathbf{x}}^*_k$ goes beyond its acceptance region, as shown in Fig.~\ref{MFalarm} with green line.
Denote the alarm time instances for fault appearance and disappearance given by MA-TCC($W$) in the online monitoring process as follows
\begin{align}
\mu^A_i(W)&\triangleq\inf\left\{k>\nu^A_{i-1}(W): T^2_k(W)>\delta^2_W\right\},\nonumber\\
\nu^A_i(W)&\triangleq\inf\left\{k>\mu^A_{i}(W): T^2_k(W)\leq\delta^2_W\right\}.
\end{align}
Having obtained a series of $\mu^A_{i}(W)$, $\nu^A_{i}(W)$, we can now use the MA-TCCs(M), which employ a bank of MA-TCCs with different window lengths simultaneously, to exclude false alarms and compensate missing alarms through the following theorems.

\begin{lemma}\label{LemRecTime}
For the MA-TCC($W$) and a given significance level $\alpha$, the disappearance of the $q$th IF is \textbf{guaranteed detectable} if and only if there exist $\{\nu^A_j(W),\mu^A_{j+1}(W)\}$ such that
\begin{align}\label{LemEquRecTime}
\nu^A_{j}(W)\leq k^{\#}_q(W)<\mu_{q+1}\leq\mu^A_{j+1}(W),
\end{align}
is guaranteed for all values of $\bar{\mathbf{x}}^*_k$ in \eqref{AssuAcceptRegion}.
\end{lemma}
\textbf{Proof.} See Appendix B. \qed

\begin{lemma}\label{LemOccTime}
For the MA-TCC($W$) and a given significance level $\alpha$, the appearance of the $q$th IF is \textbf{guaranteed detectable} if and only if there exist $\{\nu^A_j(W),\mu^A_{j+1}(W)\},\{\mu^A_i(W),\nu^A_i(W)\},j<i$ such that
\begin{align}\label{LemEquOccTime}
\nu^A_{j}(W)&\leq k^{\#}_{q-1}(W)<\mu_{q}\leq\mu^A_{j+1}(W),\nonumber\\
\mu^A_{i}(W)&\leq k^*_q(W)<\nu_q\leq k^{**}_q(W)\leq\nu^A_{i}(W),
\end{align}
are guaranteed for all values of $\bar{\mathbf{x}}^*_k$ in \eqref{AssuAcceptRegion}, where $k^{**}_q(W)$ is defined in (\ref{kxxing}).
\end{lemma}
\textbf{Proof.} See Appendix B. \qed

To facilitate reading and simplify expression, we will omit the ``$(W)$" of $\mu^A_{i}(W),\nu^A_{i}(W),\mu^d_q(W),\nu^d_q(W)$, $k^*_q(W),k^{**}_q(W),k^{\#}_{q}(W),T^2_k(W)$ in the theorems hereinafter.

\begin{lemma}\label{LemOccRecTime}
For the MA-TCC($W$) and a given significance level $\alpha$, the $q$th IF is \textbf{guaranteed detectable} if and only if there exist $\{\nu^A_j,\mu^A_{j+1}\}, \{\mu^A_i,\nu^A_i\}, \{\nu^A_l,\mu^A_{l+1}\},j<i\leq l$ such that
\begin{align}\label{EquLemOccRecTime}
\nu^A_{j}&\leq k^{\#}_{q-1}<\mu_{q}\leq\mu^A_{j+1},\nonumber\\
\mu^A_{i}&\leq k^*_q<\nu_q\leq k^{**}_q\leq\nu^A_{i},\nonumber\\
\nu^A_{l}&\leq k^{\#}_q<\mu_{q+1}\leq\mu^A_{l+1},
\end{align}
are guaranteed for all values of $\bar{\mathbf{x}}^*_k$ in \eqref{AssuAcceptRegion}.
\end{lemma}
\textbf{Proof.} Directly derived from Lemmas \ref{LemRecTime} and \ref{LemOccTime}. \qed

\begin{table*}\centering
\begin{tabular}{p{15.5cm}}\hline
{\bf Algorithm 1:} IFD based on MA-TCCs(M)\\\hline
{\bf Initialization:} Calculate $W^*$ and $W^\#$ by \eqref{EquOptWDirMag} and \eqref{EquOptWDur} with fault parameters $\{\bm{\xi}_q,f_q,\tau^r_{q-1},\tau^o_q,\tau^r_{q}\}$ or their corresponding lower bounds. Implement MA-TCCs with $W\!=\!\{W^*\!,\!\cdots\!,\!W^\#\}$ online simultaneously and denote their latest alarm time instances as $\{\nu^A_{i_W-1}(W),\mu^A_{i_W}(W),\nu^A_{i_W}(W)\}${\dag}, respectively.\\
{\bf While} $\nu^A_{i_{W^\#}}(W^\#)-\mu^A_{i_{W^\#}}(W^\#)$ satisfies \eqref{ThmEquAppDur} {\bf do}\\
\end{tabular}
\begin{tabular}{p{0.1cm}p{15.4cm}}
   {}& {\bf Missing alarms compensation:}\\
   1.& If $\exists\ W\!\in\![W^*,W^\#], i\!\in\![1,i_W]$ such that $\mu^A_{i}(W)-\nu^A_{i-1}(W)$ does not satisfy \eqref{ThmEquDispDur}, then reset $\mu^A_{i}(W)=\mu^A_{i-1}(W)$ and $\nu^A_{i-1}(W)=\nu^A_{i-2}(W)$.\\
   2.& If $\exists\ W,W'\!\in\![W^*,W^\#], i\!\in\![1,i_W], i'\!\in\![1,i_{W'}]$ such that $\left[\nu^A_{i-1}(W),\mu^A_{i}(W)\right)\cap\left[\nu^A_{i'-1}(W'),\mu^A_{i'}(W')\right)=\emptyset$, then reset $\mu^A_{i}(W)=\mu^A_{i-1}(W)$ and $\nu^A_{i-1}(W)=\nu^A_{i-2}(W)$.\\
   {}& {\bf False alarms exclusion:}\\
   3.& If $\exists\ W\!\in\![W^*,W^\#], i\!\in\![1,i_W]$ such that $\nu^A_{i}(W)-\mu^A_{i}(W)$ does not satisfy \eqref{ThmEquAppDur}, then reset $\nu^A_{i}(W)=\nu^A_{i-1}(W)$ and $\mu^A_{i}(W)=\mu^A_{i-1}(W)$.\\
   4.& If $\exists\ W,W'\!\in\![W^*,W^\#], i\!\in\![1,i_W], i'\!\in\![1,i_{W'}]$ such that $\left[\mu^A_{i}(W),\nu^A_{i}(W)\right)\cap\left[\mu^A_{i'}(W'),\nu^A_{i'}(W')\right)=\emptyset$, then reset $\nu^A_{i}(W)=\nu^A_{i-1}(W)$ and $\mu^A_{i}(W)=\mu^A_{i-1}(W)$. If $W=W^\#, i=i_{W^\#}$, then go to Initialization.\\
   {}& {\bf Inference of $\mu_{q}$ and $\nu_{q}$:}\\
   5.& For $W\in[W^*,W^\#]$ and the obtained $\mu^A_{i_W}(W),\nu^A_{i_W}(W)$, compute $\acute\mu_q(W),\grave\mu_q(W)$ and $\acute\nu_q(W),\grave\nu_q(W)$ via \eqref{ThmEquTime}. Then $[\acute\mu_q,\grave\mu_q]=[\acute\mu_q(W^*),\grave\mu_q(W^*)]\cap\cdots\cap[\acute\mu_q(W^\#),\grave\mu_q(W^\#)]$ and $[\acute\nu_q,\grave\nu_q]=[\acute\nu_q(W^*),\grave\nu_q(W^*)]\cap\cdots\cap[\acute\nu_q(W^\#),\grave\nu_q(W^\#)]$.\\\hline
\end{tabular}
\begin{tablenotes}
\item{\dag} $i_W$ is the MA-TCC($W$)'s latest alarm index.
\end{tablenotes}
\end{table*}

\begin{theorem}\label{ThmDur}
For the MA-TCC($W$) and a given significance level $\alpha$, if the $q$th IF is \textbf{guaranteed detectable}, then the corresponding alarm durations for the $(q\!-\!1)$th IF's disappearance, $q$th IF's appearance and disappearance are guaranteed to be
\begin{align}
\label{ThmEquDisp0Dur}
&\mu^A_{j+1}-\nu^A_{j}\geq\max\left\{ \tau^r_{q-1}\!-\!\nu^d_{q-1},1 \right\},\\
\label{ThmEquAppDur}
&\nu^A_{i}-\mu^A_{i}\geq\max\left\{ \tau^o_q\!+\!\nu^d_q\!-\!2\mu^d_q,W\!-\!\mu^d_q,\tau^o_q\!-\!\mu^d_q,1 \right\},\\
\label{ThmEquDispDur}
&\mu^A_{l+1}-\nu^A_{l}\geq\max\left\{ \tau^r_q\!-\!\nu^d_q,1 \right\},\quad j<i\leq l,
\end{align}
for all values of $\bar{\mathbf{x}}^*_k$ in \eqref{AssuAcceptRegion}.
\end{theorem}
\textbf{Proof.} Directly derived from Lemma \ref{LemOccRecTime}. \qed

\begin{theorem}\label{ThmAlarmsectNonempty}
For the MA-TCC and a given significance level $\alpha$, if the $q$th IF is \textbf{guaranteed detectable} with window lengths $W$ and $W'$, then
\begin{align}\label{EquAlarmsectNonempty}
\left[\nu^A_{j}(W),\mu^A_{j+1}(W)\right)\cap\left[\nu^A_{j'}(W'),\mu^A_{j'+1}(W')\right)\neq\emptyset,\\
\left[\mu^A_{i}(W),\nu^A_{i}(W)\right)\cap\left[\mu^A_{i'}(W'),\nu^A_{i'}(W')\right)\neq\emptyset,\\
\left[\nu^A_{l}(W),\mu^A_{l+1}(W)\right)\cap\left[\nu^A_{l'}(W'),\mu^A_{l'+1}(W')\right)\neq\emptyset,
\end{align}
are guaranteed for all values of $\bar{\mathbf{x}}^*_k$ in \eqref{AssuAcceptRegion}, where $j<i\leq l,j'<i'\leq l'$ are the MA-TCC($W$)'s and MA-TCC($W'$)'s alarm indices for the $q$th IF respectively.
\end{theorem}
\textbf{Proof.} If the $q$th IF is guaranteed detectable by the MA-TCC($W$) and MA-TCC($W'$), then according to Lemma \ref{LemOccRecTime}, we have $\nu^A_{j}(W)\!\leq\!\mu_q\!-\!1<\!\mu^A_{j+1}(W)$ and $\nu^A_{j'}(W')\!\leq\!\mu_q\!-\!1<\!\mu^A_{j'+1}(W')$, $\mu^A_{i}(W)\!\leq\!\nu_q\!-\!1<\!\nu^A_{i}(W)$ and $\mu^A_{i'}(W')\!\leq\!\nu_q\!-\!1\!<\!\nu^A_{i'}(W')$, $\nu^A_{l}(W)\!\leq\!\mu_{q+1}\!-\!1<\!\mu^A_{l+1}(W)$ and $\nu^A_{l'}(W')\!\leq\!\mu_{q+1}\!-\!1<\!\mu^A_{l'+1}(W')$ are guaranteed for all values of $\bar{\mathbf{x}}^*_k$ in \eqref{AssuAcceptRegion}. Thus the intersections are nonempty. \qed

\begin{remark}
When IFs are guaranteed detectable by the MA-TCC($W$), Theorem \ref{ThmDur} says if the alarm duration for fault appearance/disappearance does not satisfy \eqref{ThmEquAppDur}/\eqref{ThmEquDispDur}, then $\bar{\mathbf{x}}^*_k$ goes beyond its acceptance region and leads to false/missing alarms.
Theorem \ref{ThmAlarmsectNonempty} further says if the IFs are guaranteed detectable with several time windows, then the intersection of their detection results should not be empty set when \eqref{AssuAcceptRegion} is not violated. Therefore, Theorems \ref{ThmDur} and \ref{ThmAlarmsectNonempty} can be used to exclude false alarms and compensate missing alarms by MA-TCCs(M) in the online monitoring process, which consequently makes the proposed methods robust to process fluctuations.
For example, the time length of the yellow/green line in Fig.~\ref{MFalarm} does not satisfy \eqref{ThmEquAppDur}/\eqref{ThmEquDispDur} apparently, thus we can conclude that it is false/missing alarms. Additionally, if there is no alarm for the yellow/green line given by some other $W$, then we can also conclude that it is false/missing alarms.
\end{remark}

\subsection{Inference of $\mu_{q}$ and $\nu_{q}$}
To complete IFD task, an additional requirement is to determine IFs' appearing and disappearing time instances $\mu_{q},\nu_{q}$.
However, due to the introduction of time window and the statistical property of $T^2$ statistic, alarm delays are inevitable as given in Theorem \ref{ThmIFD}, which makes it difficult to obtain the exact time of IFs' appearance and disappearance.
Nevertheless, through the analyses for the alarm time and alarm duration caused by IFs, the range of $\mu_{q}$, $\nu_{q}$ can be determined by the following theorems.

\begin{theorem}\label{ThmTime}
For the MA-TCC($W$) and a given significance level $\alpha$, if the $q$th IF is \textbf{guaranteed detectable}, then there exist $\{\nu^A_j,\mu^A_{j+1}\}, \{\mu^A_i,\nu^A_i\}, \{\nu^A_l,\mu^A_{l+1}\},j<i\leq l$ such that
\begin{gather}
\nu^A_{j}-\nu^d_{q-1}\leq[\bm{\nu_{q-1}}]\leq\mu^A_{j+1}-W,\nonumber\\
\max\left\{ \mu^A_{i}\!-\!\mu^d_q,\nu^A_{j}\!+\!1 \right\}\leq[\bm{\mu_{q}}]\leq\min\left\{ \mu^A_{j+1},\nu^A_{i}\!-\!\mu^d_q\!-\!1 \right\},\nonumber\\
\max\left\{ \mu^A_{i}\!+\!1\!+\!\max\{\mu^d_q\!-\!\nu^d_q,0\},\nu^A_{l}\!-\!\nu^d_q \right\}\leq[\bm{\nu_{q}}]\nonumber\\
\qquad\leq\min\left\{ \nu^A_{i}+\min\{\mu^d_q\!-\!\nu^d_q,0\},\mu^A_{l+1}\!-\!W \right\},\nonumber\\
\nu^A_{l}+1\leq[\bm{\mu_{q+1}}]\leq\mu^A_{l+1},
\label{ThmEquTime}
\end{gather}
are guaranteed for all values of $\bar{\mathbf{x}}^*_k$ in \eqref{AssuAcceptRegion}. For further reference, denote $\acute\mu_q(W),\acute\nu_q(W)$ and $\grave\mu_q(W),\grave\nu_q(W)$ as the above lower and upper bounds of $\mu_q,\nu_q$, respectively.
\end{theorem}
\textbf{Proof.} According to Lemma \ref{LemOccRecTime}, if the $q$th IF is guaranteed detectable, then $\exists\{\nu^A_j,\mu^A_{j+1}\}, \{\mu^A_i,\nu^A_i\}, \{\nu^A_l,\mu^A_{l+1}\},j<i\leq l$ such that
\begin{align}\label{ThmEquTime1}
\nu^A_{j}&\leq k^{\#}_{q-1}\Leftrightarrow\nu^A_{j}-\nu^d_{q-1}\leq\nu_{q-1};\quad\nu^A_{j}<\mu_{q};\nonumber\\
k^{\#}_{q-1}&<\mu^A_{j+1}\Leftrightarrow\nu_{q-1}<\mu^A_{j+1}-\nu^d_{q-1};\quad\mu_{q}\leq\mu^A_{j+1};\nonumber\\
\mu^A_{i}&\leq{k^*_q}\Leftrightarrow\mu^A_{i}-\mu^d_q\leq\mu_{q};\quad\mu^A_{i}<\nu_q;\nonumber\\
\mu^A_{i}&\!<\!k^{**}_q\Leftrightarrow\mu^A_{i}\!+\!\mu^d_q\!-\!\nu^d_q\!<\!\nu_{q};\ k^*_q\!<\!\nu^A_{i}\Leftrightarrow\mu_{q}\!<\!\nu^A_{i}\!-\!\mu^d_q;\nonumber\\
k^{**}_q&\leq\nu^A_{i}\Leftrightarrow\nu_{q}\leq\nu^A_{i}+\mu^d_q-\nu^d_q;\quad\nu_q\leq\nu^A_{i};\nonumber\\
\nu^A_{l}&\leq k^{\#}_{q}\Leftrightarrow\nu^A_{l}-\nu^d_q\leq\nu_{q};\quad\nu^A_{l}<\mu_{q+1},\nonumber\\
k^{\#}_{q}&<\mu^A_{l+1}\Leftrightarrow\nu_{q}<\mu^A_{l+1}-\nu^d_{q};\quad\mu_{q+1}\leq\mu^A_{l+1},
\end{align}
are guaranteed for all values of $\bar{\mathbf{x}}^*_k$ in \eqref{AssuAcceptRegion}. By rewriting \eqref{ThmEquTime1} as \eqref{ThmEquTime}, the proof is complete. \qed

\begin{corollary}\label{ColInfersectNonempty}
For the MA-TCC and a given significance level $\alpha$, if the $q$th IF is \textbf{guaranteed detectable} with window lengths $W$ and $W'$, then
\begin{align}\label{EquInfersectNonempty}
\left[\acute\mu_q(W),\grave\mu_q(W)\right]\cap\left[\acute\mu_q(W'),\grave\mu_q(W')\right]\neq\emptyset,\nonumber\\
\left[\acute\nu_q(W),\grave\nu_q(W)\right]\cap\left[\acute\nu_q(W'),\grave\nu_q(W')\right]\neq\emptyset,
\end{align}
are guaranteed for all values of $\bar{\mathbf{x}}^*_k$ in \eqref{AssuAcceptRegion}.
\end{corollary}
\textbf{Proof.} If the $q$th IF is guaranteed detectable with window lengths $W$ and $W'$, then according to Theorem \ref{ThmTime}, both $\acute\mu_q(W)\!\leq\!\mu_q\!\leq\!\grave\mu_q(W)$, $\acute\mu_q(W')\!\leq\!\mu_q\!\leq\!\grave\mu_q(W')$ and $\acute\nu_q(W)\!\leq\!\nu_q\!\leq\!\grave\nu_q(W)$, $\acute\nu_q(W')\!\leq\!\nu_q\!\leq\!\grave\nu_q(W')$ are guaranteed for all values of $\bar{\mathbf{x}}^*_k$ in \eqref{AssuAcceptRegion}. Thus the intersections are nonempty. \qed

\begin{remark}
Having excluded false alarms and compensated missing alarms, we can continue to use the MA-TCCs(M) to infer $\mu_{q}$,$\nu_{q}$. According to Theorem \ref{ThmTime}, we can choose a set of window lengths which makes the IFs guaranteed detectable, and then take the intersection of all MA-TCCs' inference results as the final inference of $\mu_{q},\nu_{q}$. Corollary \ref{ColInfersectNonempty} further guarantees that this intersection is not empty when \eqref{AssuAcceptRegion} holds. Observe that the above theorems still hold if one directly replaces the fault parameters therein with their corresponding lower bounds. Thus in the case of only lower bounds are available, the proposed methodology is still effective.
Compared with using a single $W$, MA-TCCs(M) can enjoy both the improved sensitivity given by larger window lengths and the reduced inference error given by smaller window lengths. In this regard, the use of multiple $W$ can improve the IFD performance in the sense of type II error and inference error. Note that the drawback of using multiple $W$ is the increase of false alarms. Thus, methods to exclude false alarms have been developed prior to the use of multiple $W$, which guarantees a small type I error for the MA-TCCs(M).
\end{remark}

For fault detection, implement MA-TCCs with $W\!=\!\{W^*\!,\!\cdots\!,\!W^\#\}$ online simultaneously. When any of the control chart alarms, the process is considered faulty. If the alarm disappears after a while, then either a false alarm or an IF happened. Thus, the false alarms exclusion is used to discriminate IFs from false alarms. If the alarm is not removed, then we consider it as an IF and infer its appearing and disappearing time instances. Overall, the IFD algorithm based on MA-TCCs(M) is summarized as Algorithm 1. Note that the order of missing alarms compensation and false alarms exclusion may need to be adjusted according to the IF parameters.
Our main concerns have been in single mode, i.e., assuming that the process operates around a single steady-state. As for the FD problem in multi-mode, the MA-TCCs(M) can be extended by introducing a prior knowledge of the multi-mode or the multi-mode modeling techniques such as Gaussian mixture model (GMM) \cite{Yu2008Multimode} and HMM \cite{Lin2018Multimode}. In addition, when an intermittent mode or pattern is viewed as a kind of IF, the MA-TCCs(M) are applicable to detect this intermittent mode. MA-TCCs(M) can discriminate the intermittent mode from false alarms, and infer its appearing and disappearing time instances.

\section{Simulation studies}\label{SimulationSec}
\subsection{A numerical example}\label{NumSimuSub}
\begin{figure}
\begin{center}
\includegraphics[width=8.4cm]{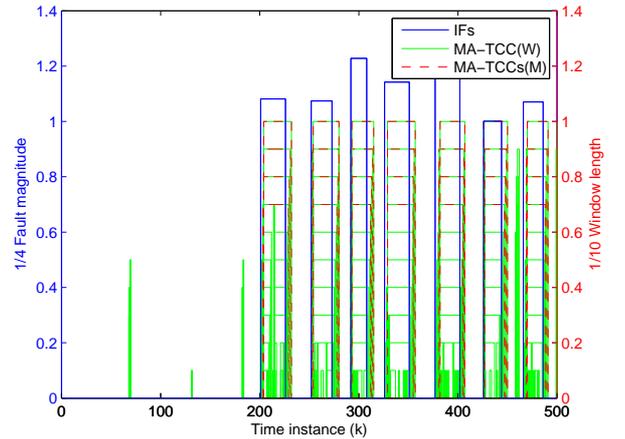}    
\caption{IFD based on MA-TCCs(M) in the numerical simulation.}
\label{MATCCsMNum}
\end{center}
\end{figure}

A simulated process model with two correlated variables is employed first. The process model under normal conditions follows a multivariate Gaussian distribution as follows
\begin{align}\label{NumericalModel}
\mathbf{x}\sim{\mathbb N}_2(\bm{\mu},\mathbf{\Sigma}), \bm{\mu}=\left[ \begin{array}{c}6\\4\end{array} \right], \mathbf{\Sigma}=\left[ \begin{array}{cc}3&{2.6}\\{2.6}&4\end{array} \right].
\end{align}
Both 5000 training samples and 500 test samples are generated according to (\ref{NumericalModel}), and intermittent process faults are subsequently introduced in the test dataset. The significance level $\alpha$ is $0.01$. The introduced IFs have an additive form as modeled by (\ref{IFModel}) with the fault direction $\bm{\xi}_q\!=\![0.2425,0.9701]^T$, the lower bound of each fault magnitude $\tilde{f}_q\!=\!4$, the lower bound of each fault active and inactive duration $\tilde{\tau}^o_q\!=\!\tilde{\tau}^r_{q}\!=\!10$. Therefore, we can conclude that the introduced IFs are guaranteed detectable by MA-TCCs with window lengths $W\in[7,10]$, according to Corollary \ref{ColIFDlow}.
The actual fault magnitude, fault active and inactive duration are all generated randomly according to their lower bounds and are shown in Fig.~\ref{MATCCsMNum} with the blue line (the left $Y$-axis shows the $1/4$ fault magnitude and the $X$-axis shows the fault active and inactive duration).

\begin{figure}
\begin{center}
\includegraphics[width=8.4cm]{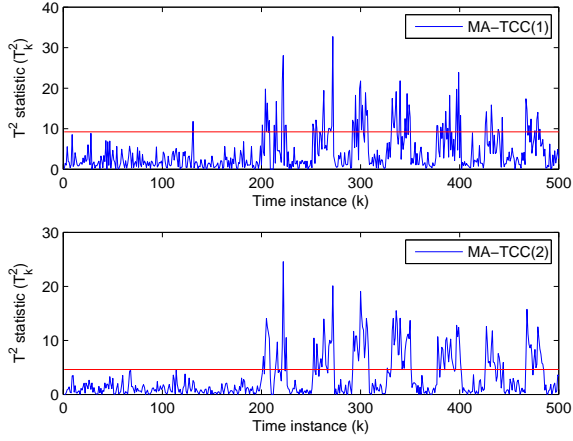}    
\caption{IFD based on the MA-TCCs with window lengths $W=1,2$ in the numerical simulation.}
\label{MATCC1a2Num}
\end{center}
\end{figure}

\begin{figure}
\begin{center}
\includegraphics[width=8.4cm]{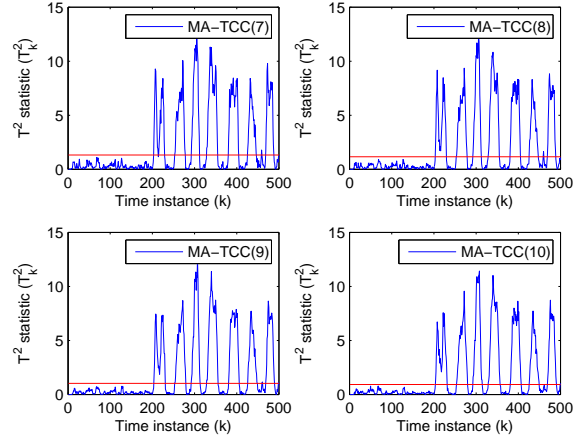}    
\caption{IFD based on the MA-TCCs with window lengths $W=7,8,9,10$ in the numerical simulation.}
\label{MATCC7a8a9a10Num}
\end{center}
\end{figure}

The MA-TCCs with window lengths $W=1,2$ of the test data have been given in Fig.~\ref{MATCC1a2Num}. It can be easily seen that the detection performance is far from satisfaction due to a large number of false and missing alarms, not to mention determining the IFs' appearing and disappearing time instances. On the contrast, the MA-TCCs with window lengths $W=7,8,9,10$ achieve better performance, which have been given in Fig.~\ref{MATCC7a8a9a10Num}.
This is because the IFs are not guaranteed detectable when $W\in[1,2]$, but are guaranteed detectable when $W\in[7,10]$.
Overall, the detailed fault detection results based on MA-TCCs with window lengths $W=1,2,\cdots,10$ have been given in Fig.~\ref{MATCCsMNum} with green lines.

On the other hand, it is noted from Fig.~\ref{MATCC7a8a9a10Num} and \ref{MATCCsMNum} that, although the IFs are guaranteed detectable when $W\in[7,10]$, false and missing alarms are still inevitable in the online monitoring process owing to the violation of condition \eqref{AssuAcceptRegion}, i.e., $\bar{\mathbf{x}}^*_k$ goes beyond its acceptance region. For example, the MA-TCC($7$) has shortly gone below its control limit before the first IF's disappears, leading to missing alarms. The MA-TCC($7$), MA-TCC($8$) and MA-TCC($9$) have shortly gone above their corresponding control limits before the seventh IF's appears, leading to false alarms.

Thanks to MA-TCCs(M) and Algorithm 1, a set of qualified window lengths are selected ($W=7,8,9,10$) and missing/false alarms can be compensated/excluded to a great extent. According to Theorems \ref{ThmDur}, \ref{ThmAlarmsectNonempty} and Algorithm 1, the alarm duration for the first IF's disappearance given by MA-TCC($7$) is too short to satisfy \eqref{ThmEquDispDur}. Moreover, neither MA-TCC($8$), MA-TCC($9$) nor MA-TCC($10$) alarms during that period. Therefore, we can conclude that they are missing alarms.
As for the false alarms given by the MA-TCC($7$), MA-TCC($8$) and MA-TCC($9$), on one hand their alarm durations for the seventh IF's appearance are too short to satisfy \eqref{ThmEquAppDur}. On the other hand, MA-TCC($10$) does not alarm during that period. Therefore, according to Theorems \ref{ThmDur}, \ref{ThmAlarmsectNonempty} and Algorithm 1, we can conclude that they are false alarms.
Overall, the detailed fault detection results after false/missing alarms exclusion/compensation based on MA-TCCs(M) have been given in Fig.~\ref{MATCCsMNum} with red lines.

\begin{table*}\centering
\small
\caption{Inferences of $\mu_q$ based on MA-TCCs with window lengths $W=7,8,9,10$ in the numerical simulation.}
\begin{tabular}{clcccccccc}\hline
{$q$} & {$\mu_q$} & $\acute\mu_q(7)$ & $\grave\mu_q(7)$ & $\acute\mu_q(8)$ & $\grave\mu_q(8)$ & $\acute\mu_q(9)$ & $\grave\mu_q(9)$ & $\acute\mu_q(10)$ & $\grave\mu_q(10)$\\\hline
1 & 201 & 197 & 203 & 197 & 204 & 197 & 204 & 196 & 204\\
2 & 252 & 248 & 254 & 246 & 253 & 246 & 253 & 246 & 254\\
3 & 292 & 287 & 293 & 286 & 293 & 285 & 292 & 285 & 293\\
4 & 326 & 323 & 329 & 322 & 329 & 322 & 329 & 321 & 329\\
5 & 377 & 375 & 381 & 374 & 381 & 374 & 381 & 374 & 382\\
6 & 426 & 420 & 426 & 419 & 426 & 420 & 427 & 418 & 426\\
7 & 466 & 463 & 469 & 462 & 469 & 462 & 469 & 462 & 470\\\hline
\end{tabular}
\label{TableMuMATCCsNum}
\end{table*}
\begin{table*}\centering
\small
\caption{Inferences of $\nu_q$ based on MA-TCCs with window lengths $W=7,8,9,10$ in the numerical simulation.}
\begin{tabular}{clcccccccc}\hline
{$q$} & {$\nu_q$} & $\acute\nu_q(7)$ & $\grave\nu_q(7)$ & $\acute\nu_q(8)$ & $\grave\nu_q(8)$ & $\acute\nu_q(9)$ & $\grave\nu_q(9)$ & $\acute\nu_q(10)$ & $\grave\nu_q(10)$\\\hline
1 & 226 & 223 & 229 & 223 & 230 & 223 & 230 & 223 & 231\\
2 & 273 & 272 & 278 & 272 & 279 & 272 & 279 & 271 & 279\\
3 & 308 & 306 & 312 & 306 & 313 & 306 & 313 & 306 & 314\\
4 & 351 & 349 & 355 & 348 & 355 & 348 & 355 & 348 & 356\\
5 & 402 & 399 & 405 & 399 & 406 & 398 & 405 & 398 & 406\\
6 & 444 & 441 & 447 & 441 & 448 & 441 & 448 & 441 & 449\\
7 & 486 & 482 & 488 & 482 & 489 & 482 & 489 & 482 & 490\\\hline
\end{tabular}
\label{TableNuMATCCsNum}
\end{table*}

Having excluded false alarms and compensated missing alarms, Algorithm 1 is further used to infer the actual appearing and disappearing time instances of IFs. Firstly, MA-TCC($7$), MA-TCC($8$), MA-TCC($9$) and MA-TCC($10$) are used to infer $\mu_q,\nu_q$ separately according to Theorem \ref{ThmTime}. Inference results have been given in Tables \ref{TableMuMATCCsNum} and \ref{TableNuMATCCsNum}, respectively.
It can be seen that the actual appearing and disappearing time instances of IFs are within the inference results.
Secondly, according to Corollary \ref{ColInfersectNonempty}, the intersection of these inference results is not empty set.
Therefore, we take the intersection of all these inference results as the final inference of $\mu_{q},\nu_{q}$, as shown in Table \ref{TableMuNuMATCCsMNum}.
Overall, it can be seen from Fig.~\ref{MATCCsMNum} and Table \ref{TableMuNuMATCCsMNum} that the developed methodology performs favorably.

\begin{table}\centering
\small
\caption{Inferences of $\mu_q, \nu_q$ based on MA-TCCs(M) in the numerical simulation.}
\begin{tabular}{clccccc}\hline
{$q$} & {$\mu_q$} & $\acute\mu_q$ & $\grave\mu_q$ & {$\nu_q$} & $\acute\nu_q$ & $\grave\nu_q$\\\hline
1 & 201 & 197 & 203 & 226 & 223 & 229\\
2 & 252 & 248 & 253 & 273 & 272 & 278\\
3 & 292 & 287 & 292 & 308 & 306 & 312\\
4 & 326 & 323 & 329 & 351 & 349 & 355\\
5 & 377 & 375 & 381 & 402 & 399 & 405\\
6 & 426 & 420 & 426 & 444 & 441 & 447\\
7 & 466 & 463 & 469 & 486 & 482 & 488\\\hline
\end{tabular}
\label{TableMuNuMATCCsMNum}
\end{table}

\subsection{The CSTR process}\label{CSTRSimuSub}
\begin{figure}
\begin{center}
\includegraphics[width=8.4cm]{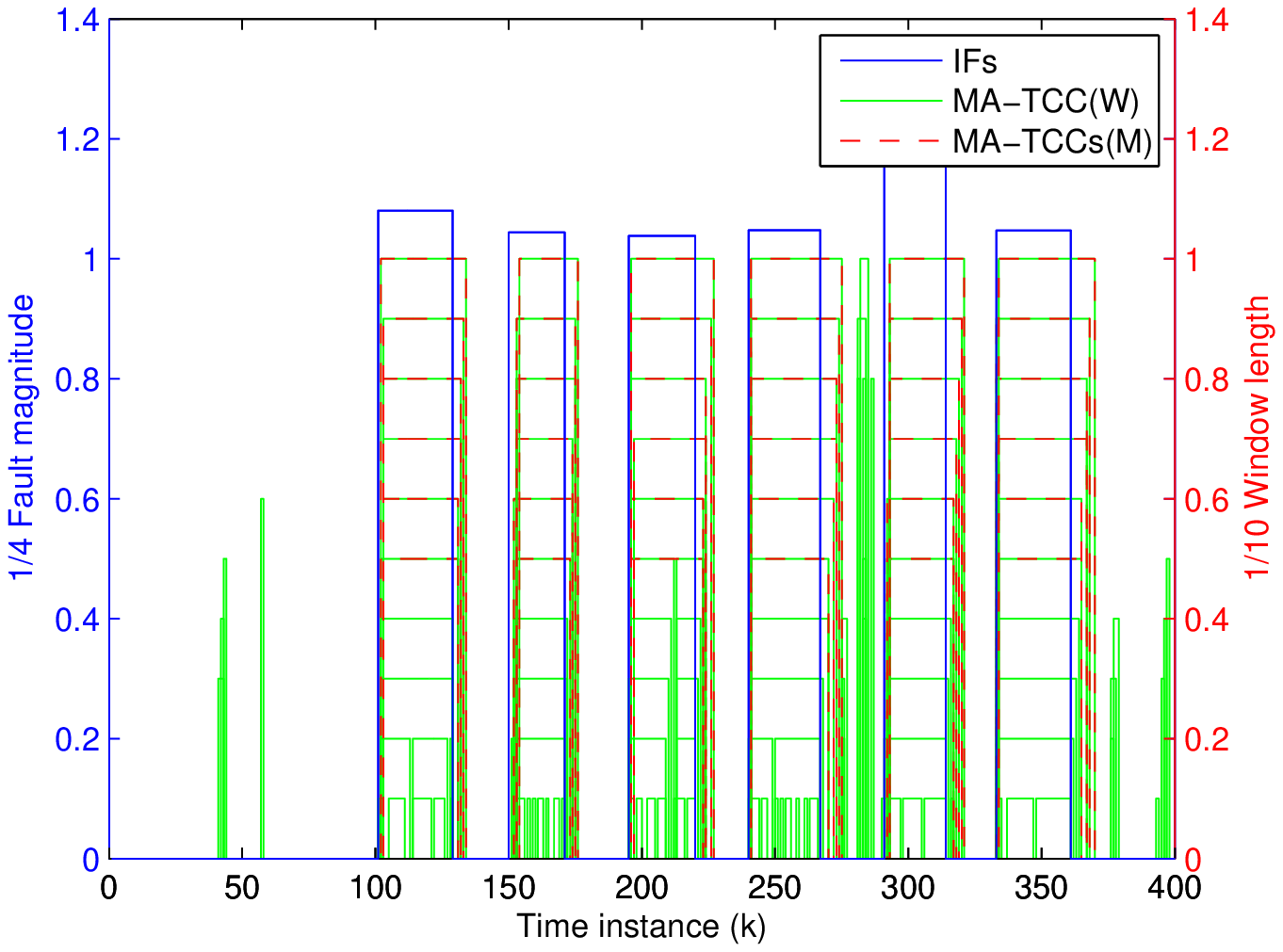}    
\caption{IFD based on MA-TCCs(M) in the CSTR process.}
\label{MATCCsMCSTR}
\end{center}
\end{figure}

In this subsection, a simulation on a continuous stirred tank reactor (CSTR) is employed to demonstrate the effectiveness of the developed methods.
The CSTR process can be described by the following differential equations
\begin{align*}
\frac{\mathrm{d}C_A}{\mathrm{d}t}&=\frac{q}{V}(C_{Af}-C_A)-k_0\exp\left(-\frac{E}{RT}\right)C_A+v_1,\\
\frac{\mathrm{d}T}{\mathrm{d}t}&=\frac{q}{V}(T_f-T)+\frac{-\Delta H}{\rho C_p}k_0\exp\left(-\frac{E}{RT}\right)C_A\\
&\qquad\qquad\qquad+\frac{UA}{V\rho C_p}(T_c-T)+v2,
\end{align*}
where $[C_A, T]^\mathrm{T}$ are controlled variables, $[T_c, q]^\mathrm{T}$ are manipulated variables, and $v_1$ and $v_2$ are independent Gaussian white noises.
Detailed descriptions of the CSTR process can be found in \cite{Li2010Reconstruction}, where the settings of the process, including system parameters and conditions as well as controller information, are also given therein.
The measured variables here are $[C_A, T, T_c, q]^\mathrm{T}$. Both 5000 fault-free training samples and 400 test samples with intermittent sensor faults are collected. The sampling interval is 30s and the significance level $\alpha$ is set as $0.01$. The faults are added to the fourth measured variable $q$ with a lower bound of each fault magnitude $\tilde{f}_q\!=\!4$, a lower bound of each fault active, and inactive duration $\tilde{\tau}^o_q\!=\!\tilde{\tau}^r_{q}\!=\!10$. The actual fault magnitudes, and fault active and inactive duration are shown in Fig.~\ref{MATCCsMCSTR} with the blue line.

\begin{figure}
\begin{center}
\includegraphics[width=8.4cm]{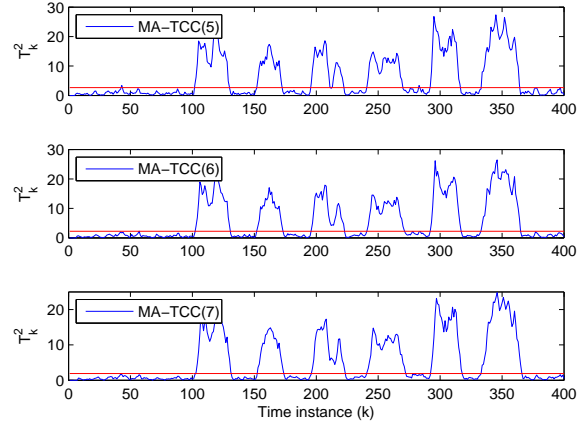}    
\caption{IFD based on the MA-TCCs with window lengths $W=5,6,7$ in the CSTR process.}
\label{MATCC5a6a7CSTR}
\end{center}
\end{figure}
\begin{figure}
\begin{center}
\includegraphics[width=8.4cm]{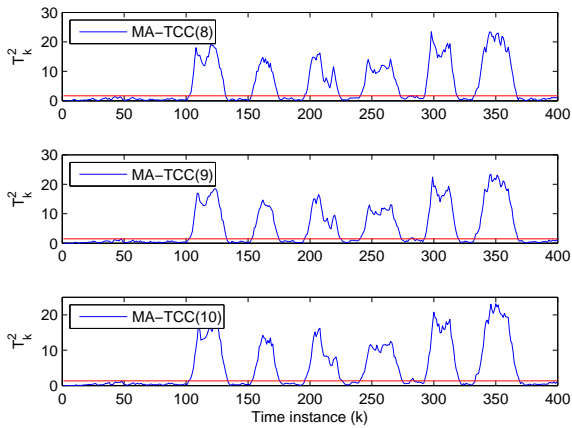}    
\caption{IFD based on the MA-TCCs with window lengths $W=8,9,10$ in the CSTR process.}
\label{MATCC8a9a10CSTR}
\end{center}
\end{figure}

\begin{figure}
\begin{center}
\includegraphics[width=8.4cm]{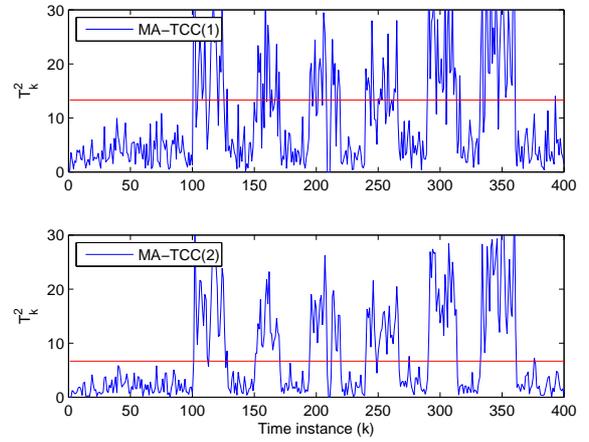}    
\caption{IFD based on the MA-TCCs with window lengths $W=1,2$ in the CSTR process.}
\label{MATCC1a2CSTR}
\end{center}
\end{figure}

According to Corollary \ref{ColIFDlow}, we can conclude that the introduced IFs are guaranteed detectable by MA-TCCs with window lengths $W\in[5,10]$, as shown in Figs.~\ref{MATCC5a6a7CSTR} and \ref{MATCC8a9a10CSTR}.
On the contrast, the MA-TCCs with window lengths $W=1,2$ do not render IFs guaranteed detectable, and consequently their detection performance is not satisfactory, as shown in Fig.~\ref{MATCC1a2CSTR}.
Overall, the detailed fault detection results based on MA-TCCs with window lengths $W=1,2,\cdots,10$ have been given in Fig.~\ref{MATCCsMCSTR} with green lines.

Similar to the above numerical example, it is noted from Figs.~\ref{MATCC5a6a7CSTR}, \ref{MATCC8a9a10CSTR} and \ref{MATCCsMCSTR} that, although the IFs are guaranteed detectable when $W\in[5,10]$, there are still false and missing alarms in the online monitoring process owing to the violation of condition \eqref{AssuAcceptRegion}. There are missing alarms within the third IF by MA-TCC($5$) as well as false alarms before the first IF and between the fourth and fifth IF.
Thus, MA-TCCs(M) have been utilized to select a set of qualified window lengths ($W=5,6,\cdots,10$) and compensate/exclude missing/false alarms, as shown in Fig.~\ref{MATCCsMCSTR} with red lines.
In addition, MA-TCCs(M) have been used to infer the actual appearing and disappearing time instances of IFs.
Firstly, MA-TCC($5$), MA-TCC($6$), ..., MA-TCC($10$) are separately used to infer $\mu_q,\nu_q$, as given in Tables \ref{TableMuMATCCsCSTR} and \ref{TableNuMATCCsCSTR}. Then, final inference of $\mu_{q},\nu_{q}$ is the intersection of all the individual inference results, which has been given in Table \ref{TableMuNuMATCCsCSTR}.
It can be seen that the actual appearing and disappearing time instances of IFs are within the inference results.
Overall, it can be seen from Fig.~\ref{MATCCsMCSTR} and Table \ref{TableMuNuMATCCsCSTR} that the developed methodology can effectively determine the added intermittent sensor faults, which confirms our theoretical results.

\begin{table*}\centering
\small
\caption{Inferences of $\mu_q$ based on MA-TCCs with window lengths $W=5,6,\cdots,10$ in the CSTR process.}
\begin{tabular}{clcccccccccccc}\hline
{$q$} & {$\mu_q$} & $\acute\mu_q(5)$ & $\grave\mu_q(5)$ & $\acute\mu_q(6)$ & $\grave\mu_q(6)$ & $\acute\mu_q(7)$ & $\grave\mu_q(7)$ & $\acute\mu_q(8)$ & $\grave\mu_q(8)$ & $\acute\mu_q(9)$ & $\grave\mu_q(9)$ & $\acute\mu_q(10)$ & $\grave\mu_q(10)$\\\hline
1 & 101 & 98 & 102 & 98 & 103 & 98 & 103 & 98 & 103 & 97 & 103 & 96 & 102\\
2 & 150 & 149 & 153 & 147 & 152 & 148 & 153 & 148 & 153 & 147 & 153 & 148 & 154\\
3 & 195 & 192 & 196 & 191 & 196 & 192 & 197 & 191 & 196 & 190 & 196 & 190 & 196\\
4 & 240 & 237 & 241 & 236 & 241 & 236 & 241 & 236 & 241 & 235 & 241 & 235 & 241\\
5 & 291 & 288 & 292 & 287 & 292 & 288 & 293 & 288 & 293 & 287 & 293 & 287 & 293\\
6 & 333 & 330 & 334 & 328 & 333 & 329 & 334 & 328 & 333 & 327 & 333 & 328 & 334\\\hline
\end{tabular}
\label{TableMuMATCCsCSTR}
\end{table*}
\begin{table*}\centering
\small
\caption{Inferences of $\nu_q$ based on MA-TCCs with window lengths $W=5,6,\cdots,10$ in the CSTR process.}
\begin{tabular}{clcccccccccccc}\hline
{$q$} & {$\nu_q$} & $\acute\nu_q(5)$ & $\grave\nu_q(5)$ & $\acute\nu_q(6)$ & $\grave\nu_q(6)$ & $\acute\nu_q(7)$ & $\grave\nu_q(7)$ & $\acute\nu_q(8)$ & $\grave\nu_q(8)$ & $\acute\nu_q(9)$ & $\grave\nu_q(9)$ & $\acute\nu_q(10)$ & $\grave\nu_q(10)$\\\hline
1 & 129 & 127 & 131 & 126 & 131 & 126 & 131 & 125 & 130 & 125 & 131 & 125 & 131\\
2 & 171 & 169 & 173 & 169 & 174 & 168 & 173 & 168 & 173 & 167 & 173 & 167 & 173\\
3 & 220 & 219 & 223 & 218 & 223 & 218 & 223 & 217 & 222 & 218 & 224 & 218 & 224\\
4 & 267 & 266 & 270 & 267 & 272 & 267 & 272 & 266 & 271 & 266 & 272 & 266 & 272\\
5 & 314 & 313 & 317 & 312 & 317 & 312 & 317 & 312 & 317 & 312 & 318 & 312 & 318\\
6 & 361 & 361 & 365 & 360 & 365 & 361 & 366 & 360 & 365 & 360 & 366 & 361 & 367\\\hline
\end{tabular}
\label{TableNuMATCCsCSTR}
\end{table*}

\begin{table}\centering
\small
\caption{Inferences of $\mu_q, \nu_q$ based on MA-TCCs(M) in the CSTR process.}
\begin{tabular}{clccccc}\hline
{$q$} & {$\mu_q$} & $\acute\mu_q$ & $\grave\mu_q$ & {$\nu_q$} & $\acute\nu_q$ & $\grave\nu_q$\\\hline
1 & 101 & 98 & 102 & 129 & 127 & 130\\
2 & 150 & 149 & 152 & 171 & 169 & 173\\
3 & 195 & 192 & 196 & 220 & 219 & 222\\
4 & 240 & 237 & 241 & 267 & 267 & 270\\
5 & 291 & 288 & 292 & 314 & 313 & 317\\
6 & 333 & 330 & 333 & 361 & 361 & 365\\\hline
\end{tabular}
\label{TableMuNuMATCCsCSTR}
\end{table}

\section{Conclusion}\label{ConclusionSec}
In this paper, moving average $T^2$ control charts with multiple window lengths (MA-TCCs(M)) have been developed for intermittent fault (IF) detection. The MA-TCCs(M) incorporate historical information through a bank of time windows and thus can improve the IFD performance. The detectability of IFs has been investigated theoretically, and choices of window lengths in different practical conditions have been discussed.
The advantage of using time window for permanent fault (PF) detection is apparent, i.e., the larger the window length $W$ is, the smaller the PF is guaranteed detectable. However, this is not the case for IFs. A window length larger than the active duration of IFs can reduce the detection capability.
Moreover, we have found that when an IF is guaranteed detectable, the corresponding alarm durations for its appearance and disappearance have lower bounds and the detection results given by a set of window lengths can be integrated. These properties have been used to exclude false alarms, compensate missing alarms and infer the fault's appearing and disappearing time instances. The numerical example and the CSTR process have demonstrated the effectiveness and applicability of our methodology.





\bibliographystyle{unsrt}
\bibliography{arXiv_IFDMATCCM-Final}

\section{Appendix A: Proofs in Sections \ref{PreliminariesSec} and \ref{DetectabilitySec}}\label{AppendixASec}
\subsection{Proof of equality \eqref{T2Conse}}
Under normal conditions, we have $\bm{\mu}_f-\bm{\mu}=\mathbf{0}$. Moreover, according to \eqref{SampleMeanCov}, we have
\begin{align*}
\sqrt{\frac{NW}{N+W}}(\bar{\mathbf{x}}^f_k-\bar{\mathbf{x}})&\sim{\mathbb N}_p(\mathbf{0},\mathbf{\Sigma}),\nonumber\\
(N-1)\mathbf{S}&\sim{\mathbb W}_p(N-1,\mathbf{\Sigma}).
\end{align*}
Note that for a multivariate normal distribution, its sample mean and sample covariance are independent. Then it follows from Lemma \ref{LemHotellingT} that
\begin{align*}
\frac{NW}{N+W}(\bar{\mathbf{x}}^f_k-\bar{\mathbf{x}})^T\mathbf{S}^{-1}(\bar{\mathbf{x}}^f_k-\bar{\mathbf{x}})\sim\frac{p(N-1)}{N-p}{\mathbb F}(p,N-p),
\end{align*}
which proves the equality \eqref{T2Conse}.

\subsection{Proof of Lemma \ref{LemIFDrec}}
According to the IF model (\ref{IFModel}), when $W\leq\tau^r_{q}$, there exists a time instance $\nu_q\leq[\bm{k^{\#}}]<\mu_{q+1}$, such that for each $k^{\#}\leq[\bm{k}]<\mu_{q+1}$, the $W$ current process samples within the time window are fault-free. Then we have $\bar{\mathbf{x}}^f_k=\bar{\mathbf{x}}^{*}_{k}$ and
\begin{align*}
T^2_k(W)=\|\mathbf{S}^{-1/2}(\bar{\mathbf{x}}^f_k-\bar{\mathbf{x}})\|^2=\|\mathbf{S}^{-1/2}(\bar{\mathbf{x}}^{*}_{k}-\bar{\mathbf{x}})\|^2.
\end{align*}
Thus, for each $k^{\#}\leq[\bm{k}]<\mu_{q+1}$, the detection statistic $T^2_k(W)\leq\delta^2_W$ is guaranteed for all values of $\bar{\mathbf{x}}^*_k$ in \eqref{AssuAcceptRegion} and for all elements of $\mathcal{P}^{q}$, and the proof of sufficiency is complete.

In the following, we will prove the necessity of \textit{Lemma \ref{LemIFDrec}} by contraposition. The contrapositive of the necessity statement is:
\textit{If $W\!>\!\tau^r_{q}$, then for any time instance $\nu_q\!\leq\![\bm{k^{\#}}]\!<\!\mu_{q+1}$, there exists a time instance $k^{\#}\!\leq\![\bm{k}]\!<\!\mu_{q+1}$, a value of $\bar{\mathbf{x}}^*_k$ in \eqref{AssuAcceptRegion} and an element of $\mathcal{P}^{q}$, making $T^2_k(W)\!>\!\delta^2_W$ valid.} It can be proven as follows.
For any given $\nu_q\!\leq\![\bm{k^{\#}}]\!<\!\mu_{q+1}$, we consider the time instance $k\!=\!\mu_{q+1}\!-\!1$ which satisfies $k^{\#}\!\leq\!k\!<\!\mu_{q+1}$.
Since $W\!>\!\tau^r_{q}$, we have $\bar{\mathbf{x}}^f_k=\bar{\mathbf{x}}^{*}_{k}+\bar{\mathbf{\Xi}}_k{\bar{\mathbf{f}}_k}$, where $\bar{\mathbf{\Xi}}_k{\bar{\mathbf{f}}_k}$ is the effect of all IFs in the time window. We further consider an element of $\mathcal{P}^{q}$ making $\|\bar{\mathbf{\Xi}}_k{\bar{\mathbf{f}}_k}\|\neq0$, and the following value of $\bar{\mathbf{x}}^*_k$ in \eqref{AssuAcceptRegion}
\begin{align*}
\mathbf{S}^{-1/2}(\bar{\mathbf{x}}^{*}_{k}-\bar{\mathbf{x}})=\frac{\mathbf{S}^{-1/2}\bar{\mathbf{\Xi}}_k{\bar{\mathbf{f}}_k}}{\|\mathbf{S}^{-1/2}\bar{\mathbf{\Xi}}_k{\bar{\mathbf{f}}_k}\|}\delta_W.
\end{align*}
Thus the detection statistic at time instance $k$ is
\begin{align*}
T^2_k(W)&=\|\mathbf{S}^{-1/2}(\bar{\mathbf{x}}^{*}_{k}-\bar{\mathbf{x}}+\bar{\mathbf{\Xi}}_k{\bar{\mathbf{f}}_k})\|^2\nonumber\\
&=\|\mathbf{S}^{-1/2}\bar{\mathbf{\Xi}}_k{\bar{\mathbf{f}}_k}\left(1+\frac{\delta_W}{\|\mathbf{S}^{-1/2}\bar{\mathbf{\Xi}}_k{\bar{\mathbf{f}}_k}\|}\right)\|^2\nonumber\\
&=\left(\delta_W+\|\mathbf{S}^{-1/2}\bar{\mathbf{\Xi}}_k{\bar{\mathbf{f}}_k}\|\right)^2>\delta^2_W,
\end{align*}
which proves the contrapositive and thus the necessity of \textit{Lemma \ref{LemIFDrec}}.

\subsection{Proof of Lemma \ref{LemIFDocc1}}
According to Lemma \ref{LemIFDrec}, when $W\leq\min\{\tau^r_{q-1},\tau^o_q\}$, the disappearance of the $(q\!-\!1)$th IF is guaranteed detectable.
According to the IF model (\ref{IFModel}), when $W\leq\min\{\tau^r_{q-1},\tau^o_q\}$, there exists a time instance $\mu_{q}\!\leq[\bm{k^*}]\!<\!\nu_{q}$, such that for each ${k^*}\leq[\bm{k}]<\nu_{q}$, the $W$ current process samples within the time window are faulty.
Thus we have $\bar{\mathbf{x}}^f_k=\bar{\mathbf{x}}^{*}_{k}+\bm{\xi}_q{f_q}$ and
\begin{align}\label{T2LowerIF1}
T^2_k(W)&=\|\mathbf{S}^{-1/2}(\bar{\mathbf{x}}^{*}_{k}-\bar{\mathbf{x}}+\bm{\xi}_q{f_q})\|^2 \nonumber\\
&\geq\left(\|\mathbf{S}^{-1/2}\bm{\xi}_q{f_q}\|-\|\mathbf{S}^{-1/2}(\bar{\mathbf{x}}^{*}_{k}-\bar{\mathbf{x}})\|\right)^2.
\end{align}
Then by following \eqref{EquIFDocc1}, \eqref{AssuAcceptRegion} and \eqref{T2LowerIF1}, we can derive that for each ${k^*}\leq[\bm{k}]<\nu_{q}$, $T^2_k(W)>\delta^2_W$ is guaranteed for all values of $\bar{\mathbf{x}}^*_k$ in \eqref{AssuAcceptRegion} and the proof of sufficiency is complete.

The contrapositive of the necessity statement is:
\textit{When $W\!\leq\!\min\{\tau^r_{q-1},\tau^o_q\}$, if $\|\mathbf{S}^{-1/2}\bm{\xi}_q{f_q}\|\!\leq\!2\delta_W$, then the disappearance of the $(q\!-\!1)$th IF is not guaranteed detectable, or for any time instance $\mu_{q}\!\leq[\bm{k^*}]\!<\!\nu_{q}$, there exists a time instance ${k^*}\leq[\bm{k}]<\nu_{q}$ and a value of $\bar{\mathbf{x}}^*_k$ in \eqref{AssuAcceptRegion}, making $T^2_k(W)\leq\delta^2_W$ valid.} It can be proven as follows.
For any given $\mu_{q}\!\leq[\bm{k^*}]\!<\!\nu_{q}$, we consider the time instance $k=\nu_{q}\!-\!1$ which satisfies ${k^*}\leq{k}<\nu_{q}$.
We further consider the following value of $\bar{\mathbf{x}}^*_k$
\begin{align}
\mathbf{S}^{-1/2}(\bar{\mathbf{x}}^{*}_{k}-\bar{\mathbf{x}})=-\mathbf{S}^{-1/2}\bm{\xi}_q{f_q}/2,
\end{align}
which is in \eqref{AssuAcceptRegion} if $\|\mathbf{S}^{-1/2}\bm{\xi}_q{f_q}\|\leq 2\delta_W$.
Moreover, since $W\leq\min\{\tau^r_{q-1},\tau^o_q\}$, we have $\bar{\mathbf{x}}^f_k=\bar{\mathbf{x}}^{*}_{k}+\bm{\xi}_q{f_q}$.
Thus the detection statistic at time instance $k$ is
\begin{align*}
T^2_k(W)=\|\mathbf{S}^{-1/2}(\bar{\mathbf{x}}^{*}_{k}\!-\!\bar{\mathbf{x}}\!+\!\bm{\xi}_q{f_q})\|^2=\|\mathbf{S}^{-1/2}\bm{\xi}_q{f_q}/2\|^2\leq\delta^2_W,
\end{align*}
which proves the contrapositive and thus the necessity of \textit{Lemma \ref{LemIFDocc1}}.

\subsection{Proof of Lemma \ref{LemIFDocc2}}
According to Lemma \ref{LemIFDrec}, when $\tau^o_q<W\leq\tau^r_{q-1}$, the disappearance of the $(q\!-\!1)$th IF is guaranteed detectable.
We consider the time instance $k^*\!=\!\nu_{q}\!-\!1$ which satisfies $\mu_{q}\!\leq{k^*}\!<\!\nu_{q}$.
According to the IF model (\ref{IFModel}), for each $k^*\!\leq\![\bm{k}]\!<\!\nu_{q}$, the latest $\tau^o_q$ current process samples within the time window $W$ are faulty and thus we have $\bar{\mathbf{x}}^f_k=\bar{\mathbf{x}}^{*}_{k}+\bm{\xi}_q{f_q}\tau^o_q/W$.
The detection statistic at time instance $k$ is then
\begin{align}\label{T2LowerIF2}
T^2_k(W)&=\|\mathbf{S}^{-1/2}(\bar{\mathbf{x}}^{*}_{k}-\bar{\mathbf{x}}+\bm{\xi}_q{f_q}\tau^o_q/W)\|^2 \nonumber\\
&\geq\left(\|\mathbf{S}^{-1/2}\bm{\xi}_q{f_q}\|\tau^o_q/W-\|\mathbf{S}^{-1/2}(\bar{\mathbf{x}}^{*}_{k}-\bar{\mathbf{x}})\|\right)^2.
\end{align}
Then by following \eqref{EquIFDocc2}, \eqref{AssuAcceptRegion} and \eqref{T2LowerIF2}, we can derive that for each ${k^*}\leq[\bm{k}]<\nu_{q}$, $T^2_k(W)>\delta^2_W$ is guaranteed for all values of $\bar{\mathbf{x}}^*_k$ in \eqref{AssuAcceptRegion} and the proof of sufficiency is complete.

The contrapositive of the necessity statement is:
\textit{When $\tau^o_q\!<\!W\!\leq\!\tau^r_{q-1}$, if $\|\mathbf{S}^{-1/2}\bm{\xi}_q{f_q}\|\tau^o_q/W\!\leq\!2\delta_W$, then the disappearance of the $(q\!-\!1)$th IF is not guaranteed detectable, or for any time instance $\mu_{q}\!\leq[\bm{k^*}]\!<\!\nu_{q}$, there exists a time instance $k^*\!\leq\![\bm{k}]\!<\!\nu_{q}$ and a value of $\bar{\mathbf{x}}^*_k$ in \eqref{AssuAcceptRegion}, making $T^2_k(W)\leq\delta^2_W$ valid.}
It can be proven as follows.
For any given $\mu_{q}\!\leq[\bm{k^*}]\!<\!\nu_{q}$, we consider the time instance $k\!=\!\nu_{q}\!-\!1$ which satisfies $k^*\!\leq\!k\!<\!\nu_{q}$.
We further consider the following value of $\bar{\mathbf{x}}^*_k$
\begin{align*}
\mathbf{S}^{-1/2}(\bar{\mathbf{x}}^{*}_{k}-\bar{\mathbf{x}})=-\mathbf{S}^{-1/2}\bm{\xi}_q{f_q}\tau^o_q/2W,
\end{align*}
which is in \eqref{AssuAcceptRegion} if $\|\mathbf{S}^{-1/2}\bm{\xi}_q{f_q}\|\tau^o_q/W\leq 2\delta_W$.
Note that at time instance ${k}$, we have $\bar{\mathbf{x}}^f_k=\bar{\mathbf{x}}^{*}_{k}+\bm{\xi}_q{f_q}\tau^o_q/W$.
Thus the detection statistic at time instance $k$ is
\begin{align*}
T^2_k(W)=\|\mathbf{S}^{-1/2}\bm{\xi}_q{f_q}\tau^o_q/2W\|^2\leq\delta^2_W,
\end{align*}
which proves the contrapositive and thus the necessity of \textit{Lemma \ref{LemIFDocc2}}.

\subsection{Proof of Theorem \ref{ThmIFD}}
The $q$th IF is guaranteed detectable if and only if $W\leq\min\{\tau^r_{q-1},\tau^r_{q}\}$ and either Lemmas \ref{LemIFDocc1} or \ref{LemIFDocc2} holds.
Note that (\ref{EquIFDocc2}) can be reformulated as
\begin{align}
\|\mathbf{S}^{-\frac{1}{2}}\bm{\xi}_q{f_q}\|>\frac{2\delta}{\tau^o_q}\sqrt{\frac{W(N+W)}{N+1}}.
\end{align}
Thus, the necessary and sufficient condition (\ref{EquIFD}) can be derived by finding the infimum of the inequalities (\ref{EquIFDocc1}) and (\ref{EquIFDocc2}), i.e., $f(W)$ with respect to $W\leq\min\{\tau^r_{q-1},\tau^r_{q}\}$, where
\begin{align*}
f(W)=&\left\{\begin{array}{ll}
2\delta\sqrt{\frac{N+W}{W(N+1)}},&{}\textrm{if}\quad W\leq\min\{\tau^r_{q-1},\tau^o_q\},\\
\frac{2\delta}{\tau^o_q}\sqrt{\frac{W(N+W)}{N+1}},&{} \textrm{if}\quad \tau^o_q<W\leq\tau^r_{q-1}.
\end{array}\right.
\end{align*}
It can be seen that when $\tau^o_q<W\leq\tau^r_{q-1}$, $f(W)$ increases as the window length $W$ increases.
When $W\leq\min\{\tau^r_{q-1},\tau^o_q\}$, $f(W)$ decreases as the window length $W$ increases.
Moreover, when $W=\tau^o_q$, we have
\begin{align}
2\delta\sqrt{\frac{N+W}{W(N+1)}}=\frac{2\delta}{\tau^o_q}\sqrt{\frac{W(N+W)}{N+1}}.
\end{align}
Thus, we can conclude that when $W\leq\min\{\tau^r_{q-1},\tau^r_{q}\}$, the infimum of $f(W)$ achieves with
\begin{align}\label{InfW}
W=\min\{\tau^r_{q-1},\tau^o_q,\tau^r_{q}\}.
\end{align}
Since $\min\{\tau^r_{q-1},\tau^o_q,\tau^r_{q}\}\leq\min\{\tau^r_{q-1},\tau^o_q\}$, by substituting (\ref{InfW}) into (\ref{EquIFDocc1}), we derive (\ref{EquIFD}).

Furthermore, if (\ref{EquIFD}) holds, we can choose the window length $W$ according to inequalities (\ref{EquIFDocc1}) and (\ref{EquIFDocc2}).
Solving inequality (\ref{EquIFDocc1}) with respect to $W\leq\min\{\tau^r_{q-1},\tau^o_q\}$ and $W\leq\min\{\tau^r_{q-1},\tau^r_{q}\}$, we obtain
\begin{align}\label{WinLengthIFDpart1}
&\frac{1}{N}\left(\frac{N\!+\!1}{4\delta^2}\|\mathbf{S}^{-\frac{1}{2}}\bm{\xi}_q{f_q}\|^2\!-\!1\right)\!>\!\frac{1}{W}\!\geq\!\frac{1}{\min\{\tau^r_{q-1}\!,\!\tau^o_q,\tau^r_{q}\}}.
\end{align}
It can be easily proven that if (\ref{EquIFD}) holds, there must be solutions for (\ref{WinLengthIFDpart1}).
Note that (\ref{EquIFDocc2}) can be reformulated as
\begin{align}\label{EquIFDocc2Ref}
\frac{(\tau^o_q)^2}{W^2}\frac{N+1}{4\delta^2}\|\mathbf{S}^{-\frac{1}{2}}\bm{\xi}_q{f_q}\|^2-\frac{N}{W}-1>0.
\end{align}
Solving the above inequality with respect to $W>0$, we have
\begin{align}\label{EquIFDocc2RefSol}
\frac{1}{W}>
\frac{ \frac{N}{2\tau^o_q}\!+\!\sqrt{\left(\frac{N}{2\tau^o_q}\right)^2\!+\!\frac{N+1}{4\delta^2}\|\mathbf{S}^{-\frac{1}{2}}\bm{\xi}_q{f_q}\|^2} }
{\tau^o_q\frac{N+1}{4\delta^2}\|\mathbf{S}^{-\frac{1}{2}}\bm{\xi}_q{f_q}\|^2}.
\end{align}
By substituting (\ref{WinLengthIFDpart1}) and (\ref{EquIFDocc2RefSol}) into (\ref{DelayIFD}), we have
\begin{align}\label{DelayIFDineq}
\sqrt{\frac{W(N+W)}{N+1}}\frac{2\delta}{\|\mathbf{S}^{-\frac{1}{2}}\bm{\xi}_q{f_q}\|}<\min\{W,\tau^o_q\}.
\end{align}
Moreover, it can be easily proven that if (\ref{EquIFD}) holds, $W=\tau^o_q$ is a solution of (\ref{EquIFDocc2Ref}).
Thus if (\ref{EquIFD}) holds, we have
\begin{align}\label{WinLengthIFDpart2Sol}
\frac{1}{\tau^o_q}>
\frac{ \frac{N}{2\tau^o_q}\!+\!\sqrt{\left(\frac{N}{2\tau^o_q}\right)^2\!+\!\frac{N+1}{4\delta^2}\|\mathbf{S}^{-\frac{1}{2}}\bm{\xi}_q{f_q}\|^2} }
{\tau^o_q\frac{N+1}{4\delta^2}\|\mathbf{S}^{-\frac{1}{2}}\bm{\xi}_q{f_q}\|^2},
\end{align}
which means there must be solutions for (\ref{WinLengthIFDpart2}).
Solving (\ref{EquIFDocc2Ref}) with respect to $\tau^o_q\!<\!W\!\leq\!\tau^r_{q-1}$ and $W\!\leq\!\min\{\tau^r_{q-1},\tau^r_{q}\}$,

(i) if $\tau^o_q<\min\{\tau^r_{q-1},\tau^r_{q}\}$, we obtain
\begin{align}\label{WinLengthIFDpart2}
&\frac{1}{\tau^o_q}>\frac{1}{W}>(\geq)\\
&\max\!\left\{ \frac{ \frac{N}{2\tau^o_q}\!+\!\sqrt{\left(\frac{N}{2\tau^o_q}\right)^2\!+\!\frac{N+1}{4\delta^2}\|\mathbf{S}^{-\frac{1}{2}}\bm{\xi}_q{f_q}\|^2} }
{\tau^o_q\frac{N+1}{4\delta^2}\|\mathbf{S}^{-\frac{1}{2}}\bm{\xi}_q{f_q}\|^2},\frac{1}{\min\{\tau^r_{q-1},\tau^r_{q}\}} \right\}\nonumber,
\end{align}

(ii) if $\tau^o_q\geq\min\{\tau^r_{q-1},\tau^r_{q}\}$, we obtain no solution.

Now we integrate (\ref{WinLengthIFDpart1}) and the above results (i), (ii) to derive (\ref{WinLengthIFD}) when (\ref{EquIFD}) holds.
When $\tau^o_q<\min\{\tau^r_{q-1},\tau^r_{q}\}$, (\ref{WinLengthIFDpart1}) can be rewritten as
\begin{align}\label{WinLengthIFDpart1rew}
&\frac{1}{N}\left(\frac{N+1}{4\delta^2}\|\mathbf{S}^{-\frac{1}{2}}\bm{\xi}_q{f_q}\|^2-1\right)>\frac{1}{W}\geq\frac{1}{\tau^o_q}.
\end{align}
Then by integrating (\ref{WinLengthIFDpart2}) and (\ref{WinLengthIFDpart1rew}), we derive (\ref{WinLengthIFD}).
When $\tau^o_q\geq\min\{\tau^r_{q-1},\tau^r_{q}\}$, by following (\ref{WinLengthIFDpart2Sol}), (\ref{WinLengthIFD}) can be rewritten as
\begin{align*}
\frac{1}{N}\left(\frac{N+1}{4\delta^2}\|\mathbf{S}^{-\frac{1}{2}}\bm{\xi}_q{f_q}\|^2-1\right)>\frac{1}{W}\geq\frac{1}{\min\{\tau^r_{q-1},\tau^r_{q}\}},
\end{align*}
which subsequently can be derived by integrating (\ref{WinLengthIFDpart1}) with the above result (ii).
Moreover, it can be seen easily from the above proof that if (\ref{EquIFD}) holds, $W=\min\{\tau^r_{q-1},\tau^o_q,\tau^r_{q}\}$ always satisfies inequality (\ref{WinLengthIFD}).

As for the alarm delay, note that for all $k^*\in\mathcal{K^*}$, the detection statistic $T^2_{k^*}(W)>\delta^2_W$ is guaranteed for all values of $\bar{\mathbf{x}}^*_{k^*}$ in \eqref{AssuAcceptRegion}, thus we have
\begin{align}\label{kxIneq}
\|\mathbf{S}^{-\frac{1}{2}}\bm{\xi}_q{f_q}\|\frac{k^*-\mu_{q}+1}{W}>2\delta\sqrt{\frac{N+W}{W(N+1)}}.
\end{align}
Similarly, for all $k^{\#}\in\mathcal{K^{\#}}$, the detection statistic $T^2_{k^\#}(W)\leq\delta^2_W$ is guaranteed for all values of $\bar{\mathbf{x}}^*_k$ in \eqref{AssuAcceptRegion} and for all elements of $\mathcal{P}^{q}$, thus we have
\begin{align}\label{kjIneq}
k^{\#}\geq\nu_{q}+W-1.
\end{align}
By considering (\ref{kxIneq}), (\ref{kjIneq}), (\ref{EquIFDrec}), and (\ref{DelayIFDineq}), we obtain (\ref{DelayIFD}) and the proof of Theorem \ref{ThmIFD} is complete.

\subsection{Proof of Theorem \ref{OptWDirMag}}
If $IF(\bm{\xi}_q, f_q, \tau^r_{q-1}, \tau^o_q, \tau^r_q)$ is guaranteed detectable, it follows from (\ref{EquIFD}) that
\begin{align*}
&\frac{1}{N}\left(\frac{N+1}{4\delta^2}\|\mathbf{S}^{-\frac{1}{2}}\bm{\xi}_q{f_q}\|^2-1\right)>\frac{1}{W^*}\geq\frac{1}{\min\{\tau^r_{q-1},\tau^o_q,\tau^r_{q}\}}.
\end{align*}
Note that $\min\{\tau^r_{q-1},\tau^o_q,\tau^r_{q}\}$ satisfies (\ref{WinLengthIFD}) according to Theorem \ref{ThmIFD2}, and consequently $W^*$ satisfies (\ref{WinLengthIFD}).
Thus, we can conclude that for all $IF(\bm{\xi}_q, f_q, \tau^r_{q-1}, \tau^o_q, \tau^r_q)\in IF(\bm{\xi}_q, f_q)$ that satisfy (\ref{EquIFD}), the MA-TCC($W^*$) can make them guaranteed detectable. The contrapositive of the above statement says for any $IF(\bm{\xi}_q, f_q, \tau^r_{q-1}, \tau^o_q, \tau^r_q)\in IF(\bm{\xi}_q, f_q)$ that is not guaranteed detectable by the MA-TCC($W^*$), it does not satisfy (\ref{EquIFD}), and thus there is no other $W$ making it guaranteed detectable. The proof of (i) is then complete.
Moreover, $W^*$ is the smallest window length that satisfies (\ref{WinLengthIFD}). Since $\mu^d_q(W)$ and $\nu^d_q(W)$ are all increasing functions of $W$, the smallest alarm delay achieves when the window length is $W^*$. The proof of (ii) is then complete.

\section{Appendix B: Proofs in Section \ref{MATCCsMSec}}\label{AppendixBSec}

\subsection{Proof of Lemma \ref{LemRecTime}}
If the disappearance of the $q$th IF is guaranteed detectable by the MA-TCC($W$), it follows from Definition \ref{DefnIFDrec} that for each $k^{\#}_q(W)\!\leq\![\bm{k}]\!<\!\mu_{q+1}$, the detection statistic $T^2_k(W)\!\leq\!\delta^2_W$ is guaranteed for all values of $\bar{\mathbf{x}}^*_k$ in \eqref{AssuAcceptRegion}.
Thus we have (\ref{LemEquRecTime}) and the proof of necessity is complete.
On the other hand, if $\exists\{\nu^A_j(W),\mu^A_{j+1}(W)\}$ such that (\ref{LemEquRecTime}) is guaranteed for all values of $\bar{\mathbf{x}}^*_k$ in \eqref{AssuAcceptRegion}, it means that $k^{\#}_q(W)\!=\!\nu_{q}\!+\!W\!-\!1\!<\!\mu_{q+1}$, namely, $W\leq\tau^r_{q}$.
Then it follows from Lemma \ref{LemIFDrec} that the disappearance of the $q$th IF is guaranteed detectable by the MA-TCC($W$), which completes the proof of sufficiency.

\subsection{Proof of Lemma \ref{LemOccTime}}
If the appearance of the $q$th IF is guaranteed detectable by the MA-TCC($W$), because of the time window, there exists a time instance $\nu_{q}\!\leq[\bm{k^{**}}]\!<\!\mu_{q+1}$ such that for each $\nu_{q}\!\leq\![\bm{k}]\!<\!k^{**}$, the detection statistic $T^2_k(W)\!>\!\delta^2_W$ is still guaranteed for all values of $\bar{\mathbf{x}}^*_k$ in \eqref{AssuAcceptRegion}. Note that all $k^{**}$ constitute a set $\mathcal{K^{**}}$ and we have
\begin{align}
&\|\mathbf{S}^{-\frac{1}{2}}\bm{\xi}_q{f_q}\|\frac{W-(k^{**}-\nu_{q})}{W}>2\delta\sqrt{\frac{N+W}{W(N+1)}}.
\end{align}
Thus,
\begin{align}
\label{kxxing}
k^{**}_q(W)\!\triangleq\!\arg\!\sup_{k^{**}\in\mathcal{K^{**}}}\!\left(k^{**}\!-\!\nu_{q}\right)\!=\!\nu_{q}\!+\!\nu^d_q(W)\!-\!\mu^d_q(W).
\end{align}
Therefore, there exist time instances $\mu_{q}\!\leq\![\bm{k^{*}_q(W)}]\!<\!\nu_q\!\leq\![\bm{k^{**}_q(W)}]\!<\!\mu_{q+1}$ such that for each $k^*_q(W)\!\leq\![\bm{k}]\!<\!k^{**}_q(W)$, the detection statistic $T^2_k(W)\!>\!\delta^2_W$ is guaranteed for all values of $\bar{\mathbf{x}}^*_k$ in \eqref{AssuAcceptRegion}.
Hence, there exist time instances $\{\mu^A_i(W),\nu^A_i(W)\}$ such that
\begin{align}\label{LemEquOccTime1}
\mu^A_{i}(W)\leq k^*_q(W)<\nu_q\leq k^{**}_q(W)\leq\nu^A_{i}(W),
\end{align}
is guaranteed for all values of $\bar{\mathbf{x}}^*_k$ in \eqref{AssuAcceptRegion}. Then by integrating (\ref{LemEquOccTime1}) and Lemma \ref{LemRecTime}, we derive (\ref{LemEquOccTime}) and the proof of necessity is complete.
On the other hand, if (\ref{LemEquOccTime}) are guaranteed for all values of $\bar{\mathbf{x}}^*_k$ in \eqref{AssuAcceptRegion}, it follows from Definition \ref{DefnIFDocc} and Lemma \ref{LemRecTime} that the appearance of the $q$th IF is guaranteed detectable by the MA-TCC($W$), which completes the proof of sufficiency.

\end{document}